\documentclass[sigconf]{acmart}

\AtBeginDocument{%
  \providecommand\BibTeX{{%
    Bib\TeX}}}

\copyrightyear{2025}
\acmYear{2025}
\setcopyright{cc}
\setcctype{by}
\acmConference[UIST '25]{The 38th Annual ACM Symposium on User Interface
Software and Technology}{September 28-October 1, 2025}{Busan, Republic of
Korea}
\acmBooktitle{The 38th Annual ACM Symposium on User Interface Software and
Technology (UIST '25), September 28-October 1, 2025, Busan, Republic of
Korea}\acmDOI{10.1145/3746059.3747797}
\acmISBN{979-8-4007-2037-6/2025/09}

\newcommand{\sysname}{Morae }
\newcommand{\sysnamens}{Morae}

\newcommand{\ipstart}[1]{\vspace{1mm}\noindent{\textbf{\textit{#1}}}}

\AtBeginDocument{%
  \providecommand\BibTeX{{%
    \normalfont B\kern-0.5em{\scshape i\kern-0.25em b}\kern-0.8em\TeX}}}

\newcommand{\amy}[1]{\textcolor{blue}{\textbf{amy:} #1}}
\newcommand {\yihao}[1]{{\color{purple}\bf{$\leftarrow$ YH: #1}\normalfont}}

\newenvironment{bluetext}{\begingroup\color{black}}{\endgroup}

\renewcommand{\amy}[1]{}
\renewcommand{\yihao}[1]{}

\usepackage{listings}
\usepackage{dblfloatfix}

\usepackage{algorithm}
\usepackage{algpseudocode}
\usepackage{amsmath} % For math equations
\usepackage{graphicx} % For including figures, if needed

\usepackage{booktabs}
\usepackage{multirow}
\usepackage{enumitem}

\usepackage[utf8]{inputenc}
\usepackage{placeins}
\usepackage{subcaption}
\usepackage{float}

\usepackage[breakable]{tcolorbox}

\begin{document}

%%
%% The "title" command has an optional parameter,
%% allowing the author to define a "short title" to be used in page headers.

\title{\sysnamens: Proactively Pausing UI Agents for User Choices}

%%
%% The "author" command and its associated commands are used to define
%% the authors and their affiliations.
%% Of note is the shared affiliation of the first two authors, and the
%% "authornote" and "authornotemark" commands
%% used to denote shared contribution to the research.
\author{Yi-Hao Peng}
\email{yihaop@cs.cmu.edu}
\affiliation{%
  \institution{Carnegie Mellon University}
  \country{}
}
\authornote{Part of this work was conducted at Adobe Research.}

\author{Dingzeyu Li}
\email{dinli@adobe.com}
\affiliation{%
  \institution{Adobe Research}
  \country{}
}

\author{Jeffrey P. Bigham}
\email{jbigham@cs.cmu.edu}
\affiliation{%
  \institution{Carnegie Mellon University}
  \country{}
}

\author{Amy Pavel}
\email{amypavel@eecs.berkeley.edu}
\affiliation{%
  % \small \institution{University of California, Berkeley} % if not using small it would become two line
  \institution{UC Berkeley}
  \country{}
}

%%
%% By default, the full list of authors will be used in the page
%% headers. Often, this list is too long, and will overlap
%% other information printed in the page headers. This command allows
%% the author to define a more concise list
%% of authors' names for this purpose.
\renewcommand{\shortauthors}{Peng et al.}

% % % for arXiv
% % Disable ACM reference format
% \settopmatter{printacmref=false, printfolios=true}

% % Remove copyright information
% \setcopyright{none}

% % Completely remove headers (conference name etc.)
% \fancyhead{}

% % Remove footnote for ACM rights text
% \renewcommand\footnotetextcopyrightpermission[1]{}

% Optional: remove ACM CCS and keywords
% \renewcommand{\shortauthors}{}
% \renewcommand{\shorttitle}{}

%%
%% The abstract is a short summary of the work to be presented in the
%% article.
\begin{abstract}

User interface (UI) agents promise to make inaccessible or complex UIs easier to access for blind and low-vision (BLV) users.
However, current UI agents typically perform tasks end-to-end without involving users in critical choices or making them aware of important contextual information, thus reducing user agency.
For example, in our field study, a BLV participant asked to buy the cheapest available sparkling water, and the agent automatically chose one from several equally priced options, without mentioning alternative products with different flavors or better ratings. 
To address this problem, we introduce {\em Morae}, a UI agent that automatically identifies decision points during task execution and pauses so that users can make choices.
Morae uses large multimodal models to interpret user queries alongside UI code and screenshots, and prompt users for clarification when there is a choice to be made.
In a study over real-world web tasks with BLV participants, Morae helped users complete more tasks and select options that better matched their preferences, as compared to baseline agents, including OpenAI Operator.
More broadly, this work exemplifies a mixed-initiative approach in which users benefit from the automation of UI agents while being able to express their preferences.

\end{abstract}

%%
%% The code below is generated by the tool at http://dl.acm.org/ccs.cfm.
%% Please copy and paste the code instead of the example below.
%%

% \begin{CCSXML}
% <ccs2012>
%  <concept>
%   <concept_id>00000000.0000000.0000000</concept_id>
%   <concept_desc>Do Not Use This Code, Generate the Correct Terms for Your Paper</concept_desc>
%   <concept_significance>500</concept_significance>
%  </concept>
%  <concept>
%   <concept_id>00000000.00000000.00000000</concept_id>
%   <concept_desc>Do Not Use This Code, Generate the Correct Terms for Your Paper</concept_desc>
%   <concept_significance>300</concept_significance>
%  </concept>
%  <concept>
%   <concept_id>00000000.00000000.00000000</concept_id>
%   <concept_desc>Do Not Use This Code, Generate the Correct Terms for Your Paper</concept_desc>
%   <concept_significance>100</concept_significance>
%  </concept>
%  <concept>
%   <concept_id>00000000.00000000.00000000</concept_id>
%   <concept_desc>Do Not Use This Code, Generate the Correct Terms for Your Paper</concept_desc>
%   <concept_significance>100</concept_significance>
%  </concept>
% </ccs2012>
% \end{CCSXML}

% \ccsdesc[500]{Do Not Use This Code~Generate the Correct Terms for Your Paper}
% \ccsdesc[300]{Do Not Use This Code~Generate the Correct Terms for Your Paper}
% \ccsdesc{Do Not Use This Code~Generate the Correct Terms for Your Paper}
% \ccsdesc[100]{Do Not Use This Code~Generate the Correct Terms for Your Paper}

%%
%% Keywords. The author(s) should pick words that accurately describe
%% the work being presented. Separate the keywords with commas.
\keywords{Agents; User Interface Agents; Proactive Agents; Human-Agent Interaction; Accessibility; Generative UI}
  
%% A "teaser" image appears between the author and affiliation
%% information and the body of the document, and typically spans the
%% page.
\begin{teaserfigure}
  \centering
  \includegraphics[width=1.0\textwidth]{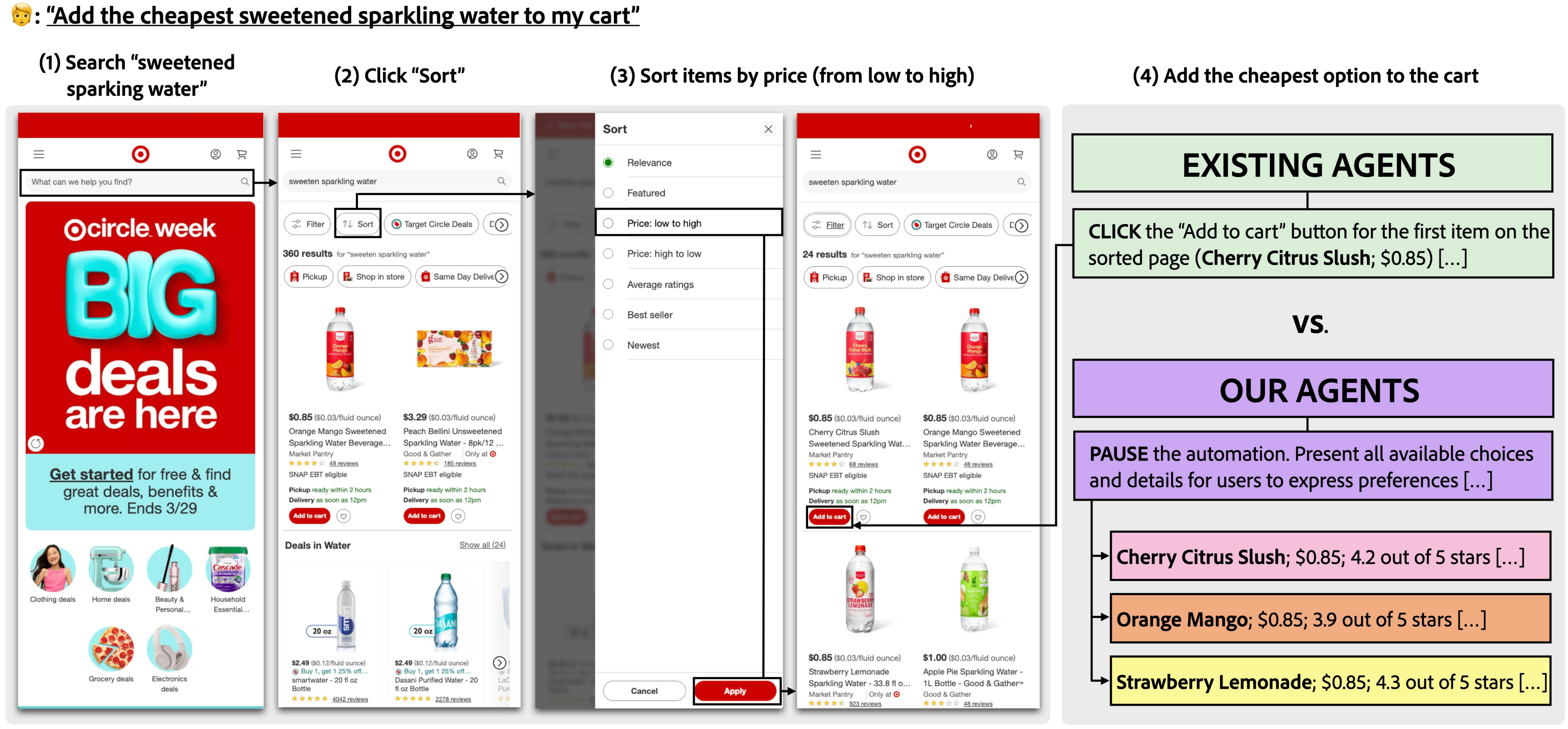}
  \caption{Morae is an accessible user interface agent that proactively pauses automation at key decision points for blind and low-vision users to make choices. For example, when asked to ``\textit{buy the cheapest sweetened sparkling water}'', existing agents would automatically choose a product even if multiple items have identical prices. Instead, Morae pauses automation and presents relevant product differences (e.g., flavors, ratings) and allows users to choose based on their preferences. By proactively detecting and pausing when user input is necessary, Morae allow users to actively express preferences during UI automation.} 
  \label{fig:teaser_pipeline}
  \Description[]{}
\end{teaserfigure}

\maketitle

\section{Introduction}

A long-standing challenge in accessible computing is to make user interfaces (UIs) both accessible and usable for people who are blind or low-vision (BLV) ~\cite{guidelines,webinsitu}. Accessibility errors with user interfaces can range from those that make access impossible to those that severely decrease usability~\cite{webinsitu}. Many BLV people use screen readers, which convey UIs in a linear form that can be slow and confusing, especially as UIs become more complex ~\cite{webinsitu,bigham2017effects}. Given the myriad challenges in making UIs accessible and usable, a promising idea is to develop agents that can work with users to navigate UIs~\cite{trailblazer,infobot}, although until recently such agents were quite limited.

Recent advances in large language and multimodal models (LLMs/LMMs) have resulted in much more capable UI agents~\cite{liu2018reinforcement, kim2024language, lai2024autowebglm, bai2024digirl} that execute complex automation tasks based on natural language instructions. For example, on a site like Target~\cite{target}, a user can instruct an agent to ``\textit{add the cheapest sweetened sparkling water to my cart.}'' The agent then finds the search bar, inputs the query, sorts the results by price, and selects the first option (Figure~\ref{fig:teaser_pipeline}). 
However, existing UI agents assume that users can visually follow the agent’s actions and understand the surrounding context. This assumption does not hold for BLV users. Consequently, even if the agent successfully adds the cheapest sweetened sparkling water, a BLV user may remain unaware that three equally priced but differently rated flavor options were available, some of which the user might have preferred (Figure~\ref{fig:teaser_pipeline}).

To understand the challenges and opportunities blind users face when interacting with UI agents, we conducted a field study with four BLV people using an open source agent to record real-world automation tasks and interaction traces. 
Our interviews with first-time users revealed enthusiasm about automating tasks such as online shopping and scheduling meetings by simply providing a few words, yet participants also noted several challenges.
Participants sometimes faced difficulties identifying the actions available in unfamiliar interfaces --- 5\% of user queries involved tasks that could not be executed (the ``not knowing what you don't know" problem" ~\cite{bigham2017effects}). Among executable tasks, 19\% were underspecified (e.g., required numerous fields/options without clear user guidance on what was required) and 13\% presented multiple options (e.g., multiple cheapest hotels or fastest flights), which led the agent to pick an arbitrary option without providing users a choice. Users also encountered challenges in tracking the agent's actions and sought additional verification to detect potential errors.

To address these challenges, we introduce \emph{Morae}\footnote{Morae, the Latin plural of mora, means  pauses or intervals. In linguistics, a mora marks the beat that measures syllable duration; this notion of incremental timing parallels the discrete pauses or steps in our agent’s action sequence during task automation.}, an accessible UI agent that supports BLV users to actively express preferences during UI automation. Morae proactively pauses at decision points whenever the user preference is unclear.
Morae leverages a large multimodal model (LMM) to interpret natural language commands and analyze UI representation (e.g., the web DOM) and screenshots. To balance effective task execution with user control, Morae introduces a mechanism called \emph{Dynamic Verification of Ambiguous Choices}, which internally verifies potential ambiguities while considering the progress of task execution. At each automation step, Morae generates and answers internal clarification questions based on relevant metadata such as the user's queries, current UI state, and previous action history. When Morae identifies critical ambiguities requiring user input, Morae explicitly pauses automation and prompts users for clarification. (Figure~\ref{fig:teaser_pipeline}). Morae also dynamically generates accessible and interactive UIs (Figure~\ref{fig:morae_ui}) that enable users to clearly specify their choices before automation continues.
To further support screen reader users, Morae provides real-time audio feedback synchronized with each agent action. Users can explicitly verify task outcomes after each automation. Morae also supports users by describing which tasks are available and how users can complete these tasks, including step-by-step instructions for screen reader-based interactions (e.g., relevant keyboard shortcuts).

We evaluated Morae through a technical assessment of its core component (i.e., the pause detection) and a user study examining its interaction experience. In our technical evaluation, we tested Morae across 256 tasks covering 8 UI types. Results indicated that Morae's approach, based on dynamic verification of ambiguity, significantly outperformed direct prompting, OpenAI Operator, and other baseline methods. Morae achieved higher task success rates and demonstrated superior performance in correctly detecting necessary pauses.
In our user study involving 10 BLV participants, Morae enabled participants to make decisions that better aligned with their preferences while maintaining more diverse choices. Users reported a stronger sense of control over the choices they made and achieved higher automation success rates compared to both the fully automated baseline agent and OpenAI Operator.

In summary, our contributions are threefold:
\textit{(i)} We release the first dataset of real-world interactions between blind users and UI agents; the dataset lays a foundation for future research on human–agent interactions that extends beyond task automation.
\textit{(ii)} We introduce \textit{Morae}, an accessible UI agent that proactively pauses at unclear decision points, prompts users to express their preferences through generative UI, and provides in-situ feedback for informed decision-making.
\textit{(iii)} A comparative study shows that Morae enables BLV users to express preferences more clearly and achieve higher task-completion rates than off-the-shelf UI agents.
\section{Related Work}
We present an accessible UI agent that proactively pauses at critical decision points, so BLV users can make choices and provide missing details.  
Our agent builds on three research strands: \textit{(i)} accessible language-based UI assistants, \textit{(ii)} language-driven interactive UI agents, and \textit{(iii)} proactive AI systems that solicit user input.

\subsection{Accessible Language-based UI Assistants}
Language-based assistants transform the way BLV users interact with digital devices. Mainstream assistants such as Siri, Alexa, and Google Assistant now perform daily tasks for users such as setting reminders or initiating phone calls via voice commands. However, these assistants primarily manage general system-level functions and rarely handle complex, application-specific interactions. Application-specific macros~\cite{rodrigues2015breaking} help automate repetitive actions within specific programs but generally lack flexibility for natural language interaction.
Recent research explores conversational UI agents for BLV users to bridge the gap between natural language commands and application-level interactions~\cite{zhong2014justspeak, ashok2015capti, pucci2023defining, phutane2023speaking, kodandaram2024enabling}. For example, JustSpeak~\cite{zhong2014justspeak} enables voice-based control of mobile applications, allowing users to operate apps through speech. Similarly, Captispeak~\cite{ashok2015capti} and ConWeb~\cite{pucci2023defining} support web browsing through natural language commands, providing more accessible online interactions. Phutane et al.~\cite{phutane2023speaking} explored conversational agents that encourage UI exploration through dialogue-based interactions. 
More recently, researchers introduced Savant~\cite{kodandaram2024enabling}, an LLM-powered desktop agent for BLV users. Savant automates desktop tasks by translating natural language commands into a series of predefined screen reader actions. This approach significantly improves efficiency and usability compared to traditional screen readers. However, like most user interface agents, Savant does not proactively seek clarification when user choices are unclear. Our study with BLV users identifies a critical limitation: users frequently give ambiguous commands without awareness, and existing agents rarely prompt users to clarify their preferences. This lack of proactive clarification leads to missed opportunities for users to express their unique preferences. We thus introduce an agent that proactively pauses at critical choice moments and allows users to specify their preferences.

\subsection{Language-driven Interactive UI Agents} 
The idea of interacting with computers using natural language began with SHRDLU, a pioneering conversational system introduced in 1968~\cite{winograd1971procedures}. The system allowed users to manipulate virtual objects in a simplified ``block world'' via English commands. SHRDLU established the foundational approach for natural language interfaces, and had a long-term impact on agent research in human-computer interaction and machine learning~\cite{li2017sugilite, wang2023enabling, srinivasan2019discovering, chen2023miwa, huang2024automatic, laput2013pixeltone,murty2024bagel,patel2024large}. Today, language-driven agents like TaxyAI~\cite{TaxyAI} and Browser Use~\cite{browser_use2024} extend natural language interfaces to automate real-world web tasks.
Recent research has also focused on using LLMs and LMMs to improve the performance and reliability of these autonomous UI agents. Kim et al.~\cite{kim2024language} demonstrated that LLMs efficiently complete tasks in simplified UI benchmarks like MiniWoB++\cite{liu2018reinforcement}, even from minimal demonstrations. Similarly, AutoWebGLM~\cite{lai2024autowebglm} further showed specialized language agents surpass GPT-4 on realistic web automation tasks in WebArena~\cite{zhou2023webarena} and Mind2Web~\cite{deng2023mind2web}. 

Recent research on UI automation explores richer interactions between users and agents. MoTIF~\cite{burns2022dataset} introduced a dataset for interactive vision-language navigation that captures ambiguous or infeasible commands. META-GUI~\citep{sun2022meta} developed multimodal conversational agents interacting directly with mobile interfaces through natural conversation, removing backend API dependency. LLM4UI~\citep{wang2023enabling} leveraged large language models to perform diverse mobile UI tasks like screen summarization, question answering, and action execution through prompting. WebLINX~\citep{lu2024weblinx} enabled conversational guidance for complex web navigation tasks through multiple dialogue turns. 
Most recently, CowPilot~\cite{huq2025cowpilot} introduced explicit buffer periods during automation. These pauses let users intervene, pause, or correct agent actions and allow task automation completed through human-agent collaboration.
Despite significant progress, existing research generally assumes that users can visually track and verify agent actions. Thus, UI automation designed for blind or low-vision (BLV) users remains largely unexplored. Our research addresses this critical gap by creating UI agents that proactively pause at unclear decision points, explicitly prompt BLV users for clarification, and provide clear audio and screen-reader-friendly feedback. With the deliberate pauses and tailored feedback mechanisms, our agent supports BLV users to actively participate and make informed and unique choices during UI automation.

\subsection{Proactive AI Agents that Solicit User Input}
Early research on mixed-initiative interfaces established continuous dialogue as a method to clarify user intent and enhance system effectiveness~\cite{horvitz1999principles}. For example, LookOut proactively extracts scheduling tasks from user emails~\cite{horvitz1999principles}, and SearchBot anticipates conversational breakdowns by proactively suggesting relevant search queries~\cite{andolina2018searchbot}. These foundational works inspired subsequent research on conversational agents that proactively initiate dialogue to enhance task performance and communication efficiency~\cite{peng2019design,marge2022spoken}. Recent research extends proactive conversation into multimodal and situated contexts. In vision-language navigation tasks, agents proactively request user clarifications to resolve ambiguous instructions, which directly improves navigation accuracy~\cite{fried2017unified, thomason2019improving, zhu2020vision}. Embodied interaction environments such as TEACh~\cite{padmakumar2022teach} and collaborative dialogue scenarios in Minecraft~\cite{narayan2019collaborative,bara2021mindcraft} illustrate proactive strategies for three-dimensional space. These agents actively clarify task-related ambiguities and align mutual understanding among users, environments, and agent goals. Similarly, in human-robot interaction, proactive techniques enable robots to detect and resolve misunderstandings during collaborative tasks. Examples include inverse semantics frameworks that request human assistance~\cite{marge2019miscommunication}, navigation strategies enhanced by retrospective curiosity-driven clarification~\cite{nguyen2019help,thomason2020vision}, and explicit frameworks addressing communication breakdown and recovery~\cite{tellex2014asking,fried2018speaker}. 
Recent works also explored more context-aware, domain-specific proactive agents. For instance, researchers have developed creation agents~\cite{hahn2024proactive} that pose clarification questions whenever multi-turn text-to-image generation encounters uncertainty, and programming agents that examine code context to propose improvements~\cite{chen2024need} or flag underspecified issues~\cite{vijayvargiya2025interactive}.
Our research builds on earlier work and introduces an agent that proactively asks BLV users to confirm choices whenever it cannot infer their preferences during UI automation. 
By offering explicit clarification steps through an agent-driven self-verification process, our approach preserves user agency and maintains task effectiveness during automated workflow.

\section{Understanding Non-Visual Use of UI Agents}

Previous research demonstrated desktop UI agents can effectively simplify interactions for BLV users compared to traditional screen readers~\cite{kodandaram2024enabling}. 
However, few research has explored the limitations BLV users encounter when adopting UI agents in-the-wild. 
We conducted a one-week field study with four BLV participants using UI agents on everyday web tasks. 
The study recorded tasks participants wanted to automate, observed agent behaviors. 
We then do a follow-up interview with each participant to collect their feedback about agent benefits and shortcomings. 
All studies were approved by the institution's IRB.

\subsection{Field Study}

\ipstart{Participants.} 
We recruited four BLV participants (three female, one male; demographic information of P1–P4 are detailed in Appendix) via mailing lists. Two participants were totally blind, and two were legally blind. All participants regularly performed web tasks in their professional roles, including accounting, software engineering, and education. None of the participants have prior experience using UI agents.

\begin{figure}
    \centering
    \includegraphics[width=1.0\linewidth]{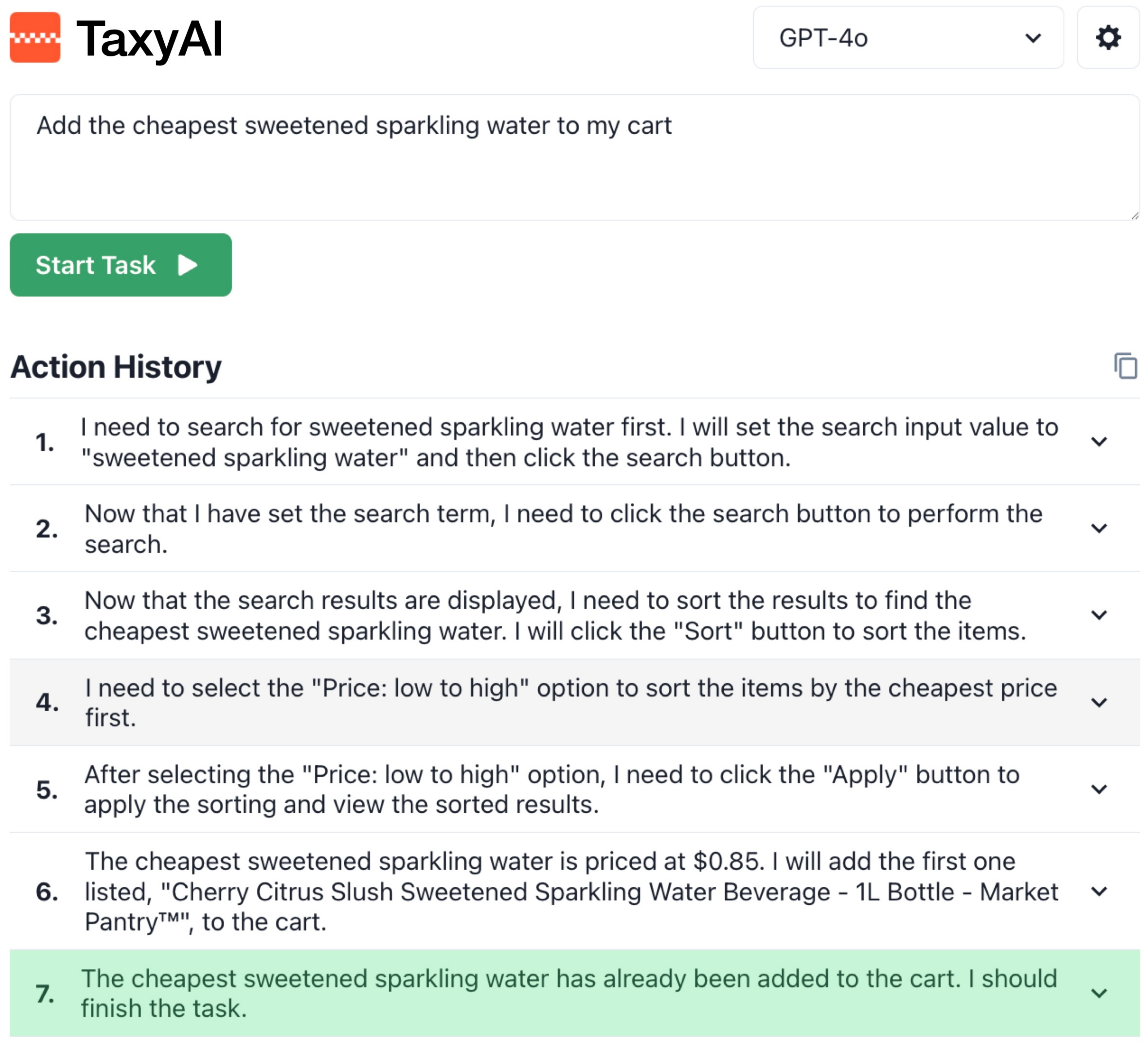}
    \caption{TaxyAI's interface features a task command input, start/pause automation button, and action history section displaying step-by-step agent reasoning. Here, TaxyAI autonomously executes the task ``\textit{add the cheapest sweetened sparkling water to my cart}'' on Target but does not ask users for their preferences even though several choices exist.}
    \label{fig:taxyai}
    \Description[]{}
\end{figure}

\ipstart{Study Apparatus.} 
For data collection, we developed an agent based on TaxyAI~\cite{TaxyAI}, an open-source, browser-based UI agent powered by an LLM (e.g., GPT-4 by default). Similar to many recent UI agents proposed by the research community~\cite{kim2024language, lai2024autowebglm, murty2024bagel, patel2024large, bai2024digirl} and industry~\cite{browser_use2024,openai_operator,deepmind_project_mariner}, TaxyAI automates web tasks by translating user commands and web UI representation (the Web DOM) into actions (e.g., clicking buttons or entering text). Due to LLM token limits, the agent first simplifies the DOM (to produce simplified DOM). It recursively traverses the DOM to remove invisible elements and redundant text nodes directly under \texttt{body}. Interactive elements, identified by accessibility attributes (e.g., \texttt{aria-label}, \texttt{role}, \texttt{name}), are retained and assigned unique IDs. Non-interactive or empty elements are removed, and single-child nodes are merged when appropriate. The resulting simplified DOM includes only the necessary interactive components to allow efficient task processing. Next, the agent pairs user commands with relevant UI elements and performs the required actions. For example, given a command like ``buy me the cheapest sparkling water to the cart'', the agent sequentially completes actions to add the item to the cart (Figure~\ref{fig:teaser_pipeline}). Users can follow the agent’s action steps through a numbered list of reasoning and interaction history in the interface (Figure~\ref{fig:taxyai}). We modified the agent so that it can take the screenshot as input to inform the actions, as well as log the model output and capture UI screenshot every step for our analyses. Additionally, we included the safety-check prompt (detailed in Appendix) as presented in prior exploration~\cite{wunderwuzzi23_system_prompts} to avoid fully automating through the high-risk tasks like checking out the orders for online shopping websites or deleting documents from cloud storage systems. 
\begin{bluetext}
We created a written document to help guide BLV users install and use browser‑based UI agents. The guide explains the concept of UI automation agents, details how to read an agent’s action history step by step, and outlines the potential outcomes and risks of UI automation with concrete example tasks.
\end{bluetext}

\ipstart{Data Processing and Annotations.}
We collected 638 unique user queries from BLV participants along with corresponding automation steps and results. Three participants used Windows (two primarily used NVDA, one used JAWS), and one participant used MacOS with VoiceOver. Two annotators independently reviewed each collected query using a structured annotation process with three stages. 
First, annotators categorized queries as \textit{(i)} valid and possible automation: meaningful queries supported by the UI, \textit{(ii)} valid but impossible automation: meaningful queries unsupported by the UI, or \textit{(iii)} invalid queries: queries without meaningful intent for automation. 
Second, for valid and possible automation queries, annotators examined screenshots, reasoning logs, and user queries to assess whether the agent’s reported ``completed'' state accurately reflected successful task completion. Annotators labeled tasks as either ``accurately completed'': fully and correctly finished, ``inaccurately completed'': the agent incorrectly indicated task completion without genuinely fulfilling the intended outcome, and ``incomplete: the agent was not able to complete the given task''.
Finally, for inaccurately completed tasks, annotators identified whether ambiguity existed such that user preference is not specified, and documented the exact automation steps where additional user input or clarification would have been necessary.
To ensure annotation consistency, annotators labeled the first 100 queries together, and achieved Cohen’s Kappa scores of 0.96 for query validity, 0.94 for task completion accuracy, and 0.82 for ambiguity and user input requirements. 
After resolving discrepancies and refining criteria, annotators independently labeled the remaining queries with evenly divided work.

\subsection{Dataset Analysis}
\ipstart{Valid and invalid user queries.} Our dataset contained 616 valid user queries and 22 invalid queries. The invalid queries typically included greetings (e.g., ``How are you doing today?'') or compliments (e.g., ``You have done a great job! Thank you!'') directed toward the UI agent. After excluding these invalid cases, we analyzed the remaining queries to characterize query types, automation outcomes, and scenarios that required additional user input (More dataset information and analysis are detailed in Appendix).

\ipstart{Types of user queries and task automation.} 
Among the valid queries, 12 queries (2\%) did not directly request task automation but instead asked about specific UI capabilities. For instance, one participant asked on Google Slides, ``Can I insert a video from my local computer onto the slide?'' Another participant queried Google Calendar capabilities by asking, ``Is it possible to automatically include a Zoom meeting link instead of Google Meet in my calendar event?'' The remaining 604 queries explicitly requested task automation.
From the 604 automation-oriented queries, the annotators categorized 28 (5\%) as impossible automation requests. These queries resulted in agent failure either because participants requested tasks on unrelated or incorrect websites (e.g., booking a hotel on Target’s website) or requested actions unsupported by the platform (e.g., generating images within Google Docs).
Participants submitted the remaining 576 feasible automation queries across 40 different platforms, representing diverse UI categories. These platforms included e-commerce websites, travel booking services (e.g., flights, hotels, restaurants), calendar and scheduling applications, productivity tools (document and slide editors), communication tools (email and messaging platforms), social media (X, Reddit), online content services (video streaming, e-learning), and cloud storage (Google Drive, Dropbox).
Within the 576 feasible queries, the agent reported reaching a ``completed'' state for 484 (84\%) tasks. The remaining 92 (16\%) tasks were ``incomplete'' due to challenges in navigating complex, multi-level interfaces, performing granular UI interactions (such as hovering or drag-and-drop), or satisfying complicated user constraints. 
\begin{bluetext}
In summary, participants sometimes issued commands that the agent could not execute because they referred to features they had not yet encountered. We view those “impossible” commands as genuine instances of users discovering unknown interface elements. We also received non-task-related inputs (such as congratulatory messages praising the agent), which we excluded from our performance analysis because they did not affect task execution.
\end{bluetext}

\ipstart{States of task completion and necessity of user input.}
For queries marked as ``completed'' by the agent, the average query length was 10.3 words, and each involved an average of 4.5 automation steps with pauses occurring at step 3.1. Among these, only 189 tasks were correctly completed. These tasks  mapped directly to one available UI option or interaction, which eliminated any ambiguity regarding user choices and preferences.
The remaining 295 queries were inaccurately labeled as completed (i.e., ``inaccurately completed'').
Out of these inaccurately completed queries, 113 involved clear execution failures. For instance, the agent mistakenly indicated task completion after adding the cheapest item to the cart without first sorting the results.
The other 182 inaccurately completed queries involved ambiguities due to incomplete user instructions or unspecified preferences. Specifically, 107 queries involved underspecified user commands. In these scenarios, essential UI fields either remained empty or kept default values because users did not specify their preferences clearly. Examples include cases where users indicated departure and arrival airports but left out critical information such as travel dates, ticket types (round-trip or one-way), or travel class (economy or business). Additionally, underspecified queries included ambiguous constraints, such as requesting the ``best chocolate'' on the website without defining criteria for ``best''. The remaining 75 inaccurately completed queries occurred when multiple valid options or UI actions matched the user's query (e.g., the example presented in Figure~\ref{fig:teaser_pipeline}), but users did not explicitly specify their preference in the original commands. For example, the agent faced multiple flavored waters with identical lowest prices when tasked with selecting the cheapest one, or encountered several available layout styles for adding page numbers when the user had not indicated a preferred style.

\begin{bluetext}
\subsection{Follow‑up Interviews}
We interviewed every participant from the field study. For each participant we chose six tasks (three that involved ambiguous choices and three that did not) from two applications the participant tried to automate (details in Appendix). Participants compared the UI agent with a screen reader alone. They reported persistent navigation difficulties with traditional screen‑reader workflows and agreed that the agent improved the task efficiency. However, they also identified gaps for existing agents, especially when ambiguous choices arose and the agent failed to ask clarifying questions.

\ipstart{How Do UI Agents Improve Task Execution?}
None of the participants had used a UI agent before the study. After hands‑on experience, every participant stated that language‑based automation boosted both accessibility and efficiency. P2 remarked, ``\textit{The agent automates repetitive operations and lets me decide what I want; it handles the rest.}'' Participants completed 40\% of the tasks with the agent versus 25\% with only a screen reader and worked roughly five times faster (42s per task versus 217s). They valued the agent's ability to process complex or repetitive actions, which allowed them to focus on key decisions.

\ipstart{What Challenges Remain?}
During the study the agent often acted without clarifying user intent. Interviews revealed that in 95\% of these situations participants never realised that multiple valid options existed. P4 noted, ``\textit{Without this interview I wouldn't know there were several choices at the same price. I need prompts that surface differences and invite my input.}''
Participants also struggled to specify complex preferences and to monitor progress. P3 said, ``\textit{When many fields appear, I lose track of the required information. A clear scaffold that lists pending items and allows review would help.}''
Participants further highlighted limited real‑time feedback. The visual action‑history log offered some traceability, yet it seldom conveyed success or failure accurately. Several tasks appeared complete even when the agent had failed. Participants asked for audible or textual cues during execution and explicit confirmation at the end. Finally, participants wanted guidance on manual task completion to ensure independence when automation falls short.

\ipstart{Design Opportunities for Accessible UI Agents.}
Insights from the interviews suggest five design directions that address the reported challenges and keep BLV users in control:
\begin{itemize}[leftmargin=*]
\item \textbf{D1: Active choice}. Pause at decision points and describe key differences among alternatives. For example, list the cheapest items side‑by‑side so users can select before the agent proceeds.
\item \textbf{D2: Clear preference input}. Provide structured widgets such as drop‑downs or number pickers that capture details text commands might miss, enabling refinement without re‑typing the full query.
\item \textbf{D3: Real‑time feedback}. Announce ongoing actions through audio or concise status text, instead of waiting until completion, to keep users oriented during lengthy or multi‑step tasks.
\item \textbf{D4: Result verification}. Prompt users to confirm outcomes—for example, “Page numbers added---accept or undo?”---and offer quick links to inspect changes, which prevents silent failures.
\item \textbf{D5: Task literacy}. Explain the UI’s available actions and outline manual workflows so users can complete tasks with a screen reader when automation tools are unavailable.
\end{itemize}
\end{bluetext}

\section{Morae}
Informed by field studies and interviews, we developed Morae, a UI agent supporting BLV users' active engagement in UI automation. Morae addresses identified needs by providing: \textit{(i)} dynamic ambiguity verification for user clarification of choices (D1), \textit{(ii)} interactive preference input methods (D2), \textit{(iii)} in-situ feedback on agent actions and outcomes (D3, D4), and \textit{(iv)} a query mechanism for available actions and screen reader task guidance (D5). 
To build Morae, we extended TaxyAI, an existing UI agent used in our earlier studies. We chose TaxyAI because it runs as a Chrome extension, requires minimal user system dependencies, and is straightforward to deploy. The agent operates through the following steps: At each interaction step, Morae captures the UI state as a reduced-DOM tree (Section 3.1) accompanied by a screenshot. Then, an LMM (GPT-4o) translates the user's natural-language commands, UI observations, and previous action history into executable actions.

\subsection{Dynamic Verification of Ambiguous Choices}
We aimed to ensure that Morae pauses at appropriate times. To identify suitable pause conditions, we analyzed our validation dataset. This dataset comprises 30\% (54 out of 182 tasks) of all annotated tasks marked as "unclear user choice or preference" during our field study. Based on this analysis, we introduced an ambiguity-aware algorithmic framework. The framework dynamically determines optimal pause points by utilizing a self-ask-then-answer verification strategy (specified prompts and model parameters are detailed in Appendix):

~\\\noindent At each interaction step \( i \), given user command \( Q \), observed UI state \( V(i) \), and executed action history \( H(i) \), the agent follows three stages to determine the next actions:

~\\\noindent\textbf{Stage 1: Critical Actions.} 
Our agent aims to balance autonomous task execution with timely intervention. We define \textit{critical actions} as those involve user-defined preferences or reveal essential UI details required for meaningful ambiguity verification. Actions at planned step \( i \) are divided into critical (\( P_c(i) \)) and non-critical (\( P_n(i) \)):
\[
P(i)=P_c(i)\cup P_n(i),\quad P_c(i)\cap P_n(i)=\emptyset.
\]
Critical actions are prioritized to ensure accurate ambiguity assessment such that the agent only pauses when necessary.

~\\\noindent\textbf{Stage 2: Ambiguity Verification.} For each step, the agent formulates prioritized ambiguity-verification questions to uncover and assess various aspects of potential ambiguity cases based on user command \( Q \), observed UI state \( V(i) \), and executed action history \( H(i) \). The example verification aspects include 
selection ambiguity, where multiple UI options or actions meet the user's criteria, and specification ambiguity, where user's command is incomplete or involving ambiguous requirements.
The agent answers each verification question with one of four responses (yes, no, unanswerable and proceed, not important and proceed). An ambiguity indicator \( A(i) \) is defined:
\[
A(i)=
\begin{cases}
1,~\text{if any question returns ``yes''},\\
0,~\text{otherwise}.
\end{cases}
\]
The agent also explicitly evaluates the sufficiency of observed UI details using indicator \( I(i) \):
\[
I(i)=
\begin{cases}
1,~\text{if agent's recorded details enable informed user decisions},\\
0,~\text{otherwise}.
\end{cases}
\]
\\
\noindent\textbf{Stage 3: Decision Function.} Finally, the decision function \( D(i) \) combines task execution and ambiguity verification, explicitly determining preference-elicited pauses:
\[
D(i)=
\begin{cases}
\text{Execute critical actions},&\text{if incomplete},\\
\text{Pause for clarification},&\text{if }A(i)=1,I(i)=1,\\
\text{Gather more UI details},&\text{if }A(i)=1,I(i)=0,\\
\text{Proceed with next planned actions},&\text{if }A(i)=0.
\end{cases}
\]
Our algorithmic approach balances effective task execution with the need to detect and clarify unclear user choices or preferences. The agent dynamically verifies whether pausing is necessary at each interaction step.
\begin{bluetext}
Our prompt design strategy builds general templates from field‑study observations. These templates adapt to varied scenarios yet require few task‑specific changes. The approach ensures broad applicability and enables straightforward reuse across diverse tasks.
\end{bluetext}

\subsection{Generative UI for Capturing User Choices}
\begin{figure}[htp]
    \centering
    \includegraphics[width=1.0\linewidth]{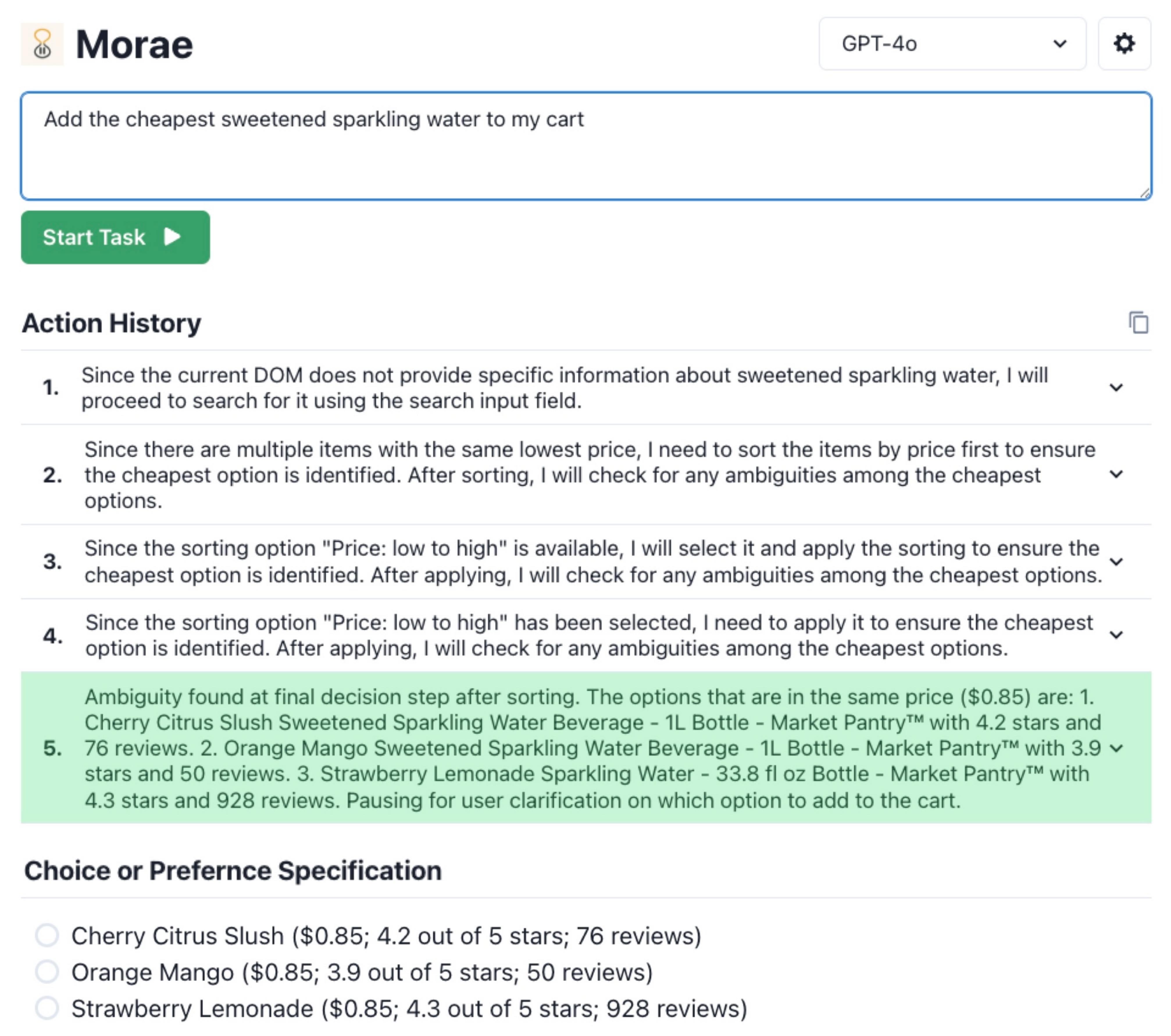}
    \caption{Morae's interface features a task command input field, a start/pause automation button, an action history panel that displays the agent’s step‑by‑step reasoning, and dynamically generated UI elements for preference selection. In this scenario, Morae autonomously executes the command ``\textit{add the cheapest sweetened sparkling water to my cart}'' on Target. At a decision point that needs user input, Morae pauses, presents relevant information, and offers accessible interactive controls so BLV users can state their choices in a structured manner.}
    \label{fig:morae_ui}
    \Description[]{}
\end{figure}

When Morae detects ambiguity in user choices during task automation, the agent proactively pauses the process and prompts users to clarify their preferences. Users can respond by text, but managing multiple decisions simultaneously can be challenging. For example, if a user asks the agent to find the fastest flight between two cities but omits travel dates, ticket type (one-way or round-trip), and travel class (economy or business), the agent pauses. It explicitly lists missing details and informs users of preset defaults.
To simplify preference specification, our agent dynamically generates a UI that clearly displays available options and thus enables users to easily specify their preferences (Figure~\ref{fig:morae_ui}). To achieve dynamic UI generation, the agent first identifies required decision properties and fields by analyzing the user's command \( Q \) alongside the current UI state \( V(i) \). Based on the identified elements, the agent then creates appropriate UI components on the agent interface, such as radio buttons for single selections or text fields for numeric or text inputs (e.g., meeting time or topics). Each decision aspect is clearly labeled with proper header levels for screen reader navigation.
Users can then optionally interact with the generated UI to clarify preferences, and their inputs guide the agent's subsequent actions.

\subsection{In-situ Feedback for Screen Reader Users}
To support active decision-making, we incorporate multiple forms of in-situ feedback that help BLV users remain involved throughout UI automation.
Drawing from feedback collected in our prior studies, our agent provides real-time auditory feedback corresponding directly to agent actions and states. For example, a distinct clicking sound plays when the agent selects buttons or links, while a typing sound indicates inputting text into fields. If the agent encounters ambiguity and requires user clarification, it emits a unique prompting tone. Conversely, upon successfully completing an action, the agent provides a distinct confirmation sound. Furthermore, users have the option to request additional visual verification after completing tasks. As GPT-4o already drives the primary automation actions, we integrated Gemini-2.0-flash, another LMM, to perform supplementary visual verification.
In addition, our agent actively supports blind users' exploration and learning of interfaces by responding to user-posed questions about UI. The agent additionally classifies user queries into two categories---questions about the UI or commands for automation. To evaluate the accuracy of this classification approach, we compared interface-related question samples gathered during our field study against an equal number of randomly selected automation-related queries. This evaluation demonstrated the existing model can achieve accuracy in distinguishing between these categories, achieving 96\% accuracy.
When responding to interface-specific questions (i.e., what tasks I can do with the UI and how to do specific tasks), the agent combines external knowledge encoded in the LMM with the current UI state observations. To provide personalized actionable guidance, we include the user's screen reader choice gathered from the survey directly within the model's prompt. For example, when a user asks, \emph{``How do I find my recent emails in Gmail?"}, the agent provides detailed, screen reader-specific instructions. These instructions clearly outline NVDA shortcuts, such as pressing \texttt{NVDA + F7} to open the links list, selecting "Inbox," and navigating emails with arrow keys, thereby facilitating precise and accessible interaction.

\section{Technical Evaluation}
We evaluated the core computational component of our agent, which determines \textit{if} and \textit{when} the agent should pause to request user input to make choices or specify preferences. Our evaluation benchmarks our method against several baseline methods derived from variations of our method and the OpenAI Operator. We conducted these comparisons on a subset of annotated tasks from our field study.

\subsection{Methods} 
We constructed the test set from two groups of tasks across 20 different application platforms. The first group included tasks identified during our field study as having ambiguous user choices or unclear preferences. From this group, we selected 128 tasks (70\%), sampling evenly from each of the eight application categories. The second group consisted of tasks that users marked as completed without any ambiguous preferences. Similarly, we randomly selected another 128 tasks from this group, ensuring equal representation from the same eight categories.
We ensured the reproducibility of all selected tasks without needing access to personal user accounts or data. For instance, tasks conducted on platforms like Google Drive were recreated using our own accounts to generate identical interaction traces. This approach allowed us to accurately produce ground-truth labels and evaluate the agent in our online setup.
For each task in the test set, we manually recorded and verified two interaction paths: \textit{(i)} a complete sequence of steps required for successful task completion without pausing, and
\textit{(ii)} a sequence of steps leading up to the point when the agent paused to request user input.
On average, the agent required 5.4 interaction steps per task, with pauses typically occurring at step 3.8.

Unlike conventional offline evaluations, which statically assess agents based on fixed queries, UI states, and interaction histories, we employed an online evaluation approach~\cite{pan2024webcanvas,xue2025illusion}. Our online evaluation captures the dynamics between the agent's execution steps and pausing decisions, agent behavior randomness, and variations in observed UI states.
We evaluated each agent condition based on two performance criteria:
\textit{(i)} \textbf{Task success rate:}  
The agent must either fully execute all annotated ground-truth steps for tasks requiring no pauses or correctly pause exactly at the annotated step for tasks needing user input.
\textit{(ii)} \textbf{Pausing performance:}  
We categorized agent pausing decisions at the task level into four possible outcomes:
\begin{itemize}[leftmargin=*]
    \item \textbf{True Positive (TP):} The agent correctly pauses exactly at the annotated step in tasks requiring a pause.
    \item \textbf{False Positive (FP):} The agent pauses incorrectly or unnecessarily. False positives occur when the agent pauses prematurely (before the annotated step) in pause-required tasks, or at any step in tasks requiring no pause.
    \item \textbf{False Negative (FN):} The agent fails to pause at the explicitly annotated pause step. This scenario includes cases when the agent continues without any pausing or pauses at a later step after missing the correct annotated step.
    \item \textbf{True Negative (TN):} The agent successfully completes all annotated steps without pausing in tasks requiring no pauses. We exclude tasks from TN calculations if the agent neither pauses nor successfully completes the task.
\end{itemize}

\begin{figure}[htbp]
    \centering
    \includegraphics[width=1.0\linewidth]{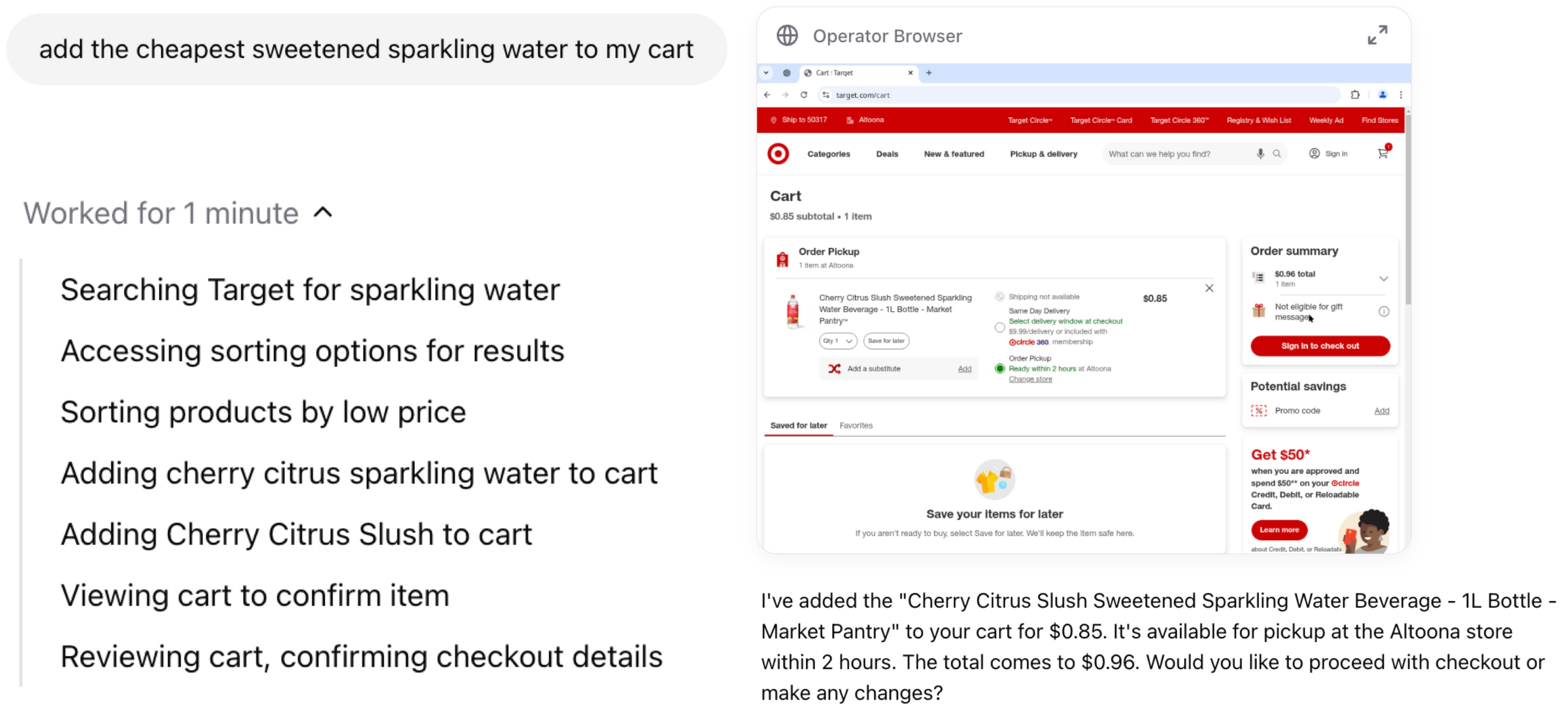}
    \caption{The UI of OpenAI's Operator Agent (zoomed view). Users assign tasks via chat, monitor browser automation through live streaming and text updates, and manually intervene at any time. The example illustrates Operator autonomously executing the command ``add the cheapest sweetened sparkling water to my cart'' on Target without pausing to ask for the user's choice.
    }
    \label{fig:operator}
    \Description[]{}
\end{figure}

\begin{table*}[htbp]
\centering
\caption{Task completion rate and pausing performance for each method we benchmark in our technical evaluation. }
\label{tab:tech_eval}
\scalebox{0.95}{%
\begin{tabular}{@{}lccc|ccc@{}}
\toprule
\multirow{2}{*}{\textbf{Method}} & \multicolumn{3}{c|}{\textbf{Task Success Rate}} & \multicolumn{3}{c}{\textbf{Pausing Performance}} \\ 
\cmidrule(lr){2-4} \cmidrule(l){5-7}
                                 & Pause Required & Pause Not Required & \textbf{Overall} & Precision & Recall & \textbf{F1-score} \\ 
\midrule
Prompting                        & 18.0\%          & 33.1\%             & 25.5\%           & \textbf{60.0\%}  & 18.6\% & 28.4\%   \\
Verifying-First-Step             & 32.3\%          & 25.0\%             & 28.6\%           & 30.2\%           & 40.8\% & 34.8\%   \\
Verifying-Per-Step               & 41.1\%          & 27.9\%             & 34.5\%           & 32.6\%           & 57.9\% & 41.7\%   \\
OpenAI Operator                         & 50.8\%          & \textbf{55.5\%}    & 53.1\%           & 58.7\%           & 59.1\% & 59.0\%   \\
\midrule
% \rowcolor[HTML]{EFEFEF}
\textbf{Verifying-Per-Step-with-Planning (Ours)} & \textbf{65.6\%} & 44.8\% & \textbf{55.2\%} & 59.7\% & \textbf{69.8\%} & \textbf{64.4\%} \\ 
\bottomrule
\end{tabular}%
}
\end{table*}

We compared several variants of our method as baselines. All these agent variations including our method  are built on TaxyAI and utilized GPT-4o as their LMM backbone. Their primary differences were the instructions or mechanisms used to trigger pauses (detailed prompts in the Appendix). Additionally, we included OpenAI Operator as a strong baseline. OpenAI Operator~\cite{openai_operator} was explicitly trained for UI operations and optimized for user interactivity and safety. 
The methods we compared were: 
\begin{itemize}[leftmargin=*]
    \item \textbf{Prompting}: The agent received explicit instruction to pause whenever user choices or preferences appeared unclear, supplemented by three descriptive examples. 
    \item \textbf{Verifying-First-Step}: The agent generated ambiguity verification questions at the start of task execution. Questions were resampled 3 times and only picked top-5 as the verification question set. Questions relied solely on the initial user query and UI observation and remained constant throughout task execution steps.
    \item \textbf{Verifying-Per-Step}: At every execution step, the agent generated new ambiguity verification questions based on the current user query, observed UI states, and prior action history. These verification questions differed at each step.
    \item \textbf{Verifying-Per-Step-with-Planning (Our Method)}: At each step, our method dynamically generated ambiguity verification questions based on user's query, observed UI and prior actions. Furthermore, the agent incorporated planning considerations to balance task automation progress with proactive user input solicitation when user preferences appeared unclear. 
    \item \textbf{OpenAI Operator}~\cite{openai_operator}: A state-of-the-art UI agent~\cite{xue2025illusion}, fine-tuned explicitly on UI operations. We used the official client interface (Figure~\ref{fig:operator}) to execute all tasks. 
\end{itemize}
After each task execution completed, we closed the browser and reopened the browser for the next task. Due to the nondeterministic nature of the agent outputs (even when setting temperature for the underlying model to be 0), we ran each task query three times for each method and compare the outcomes with the ground-truth annotations. We report the averaged results as the final task performance for each condition.

\subsection{Results} 
Our approach achieved the highest average task success rate of 55.2\%, outperforming the baseline Operator by 2.1\%. The main improvement came from tasks that required pauses, where our approach increased success rates from 50.8\% (Operator) to 65.6\%. We also observed a significant recall improvement of 10.7\% over Operator. The self-verification process primarily contributed to this gain by improving sensitivity to unclear user choices, such as distinguishing multiple calendar entries labeled "Tuesday" for scheduling a meeting. 
Compared to other variants of our method, our approach consistently demonstrated higher success rates across both pause-required and non-pause tasks. For tasks without pauses, the agent effectively planned and prioritized critical actions, continuously monitored task progress, and anticipated potential needs for user input. For tasks requiring pauses, the self-verification method more accurately determined the correct moment to request user input than simpler prompting methods. 
Additionally, our method balanced proactive detection of pauses with continuous task execution to avoid unnecessary pauses. Alternative variants relying only on verification at initial or intermediate stages frequently paused prematurely to resolve ambiguities without fulfilling task requirements or obtaining sufficient context for user decision.
While our approach still paused unnecessarily (precision errors) or missed opportunities to pause (recall errors), it demonstrated overall greater robustness compared to other methods in automating tasks while accurately identifying necessary pauses to clarify user preferences (More failure cases are detailed in Appendix).

\section{User Evaluation}

We conducted a user evaluation to assess the usability of Morae by comparing it with two baseline agents: the latest TaxyAI (with gpt-4o as the LMM backbone) and OpenAI Operator. Each participant performed nine tasks across three applications using all 3 agents.

\subsection{Method}

\ipstart{Participants.}
We recruited 10 BLV participants aged between 28 and 55 (U1–U10; 4 female, 6 male; details in Appendix) through online mailing lists. The evaluation involved a 2-hour remote Zoom session. Participants had diverse professional backgrounds including students, customer support specialists, accessibility consultants, and software engineers. All participants have extensive experience with screen readers, general web browsing, and AI-based accessibility tools such as ChatGPT, BeMyAI, and SeeingAI. 
None had previously participated in our formative study nor used the UI agent before. 
Participants received compensation of \$100 per hour, and the institution's IRB approved the study .

\ipstart{Procedure.}
The study began with demographic data collection and an overview highlighting core features of each agent. Before the study, participants reviewed text documentation for Morae, TaxyAI, and Operator and know how to use each agent independently.
Participants completed three tasks per application across three popular websites:
\textit{(i)} Target~\cite{target}: ``\textit{adding the highest-rated beer to the cart}``, ``\textit{finding details of the latest video game deal of specific brand}``, and ``\textit{buying the best seller short}''
\textit{(ii)} Google Calendar~\cite{calendar}: ``\textit{turning off the notification}'', ``\textit{adding a new event}'', and ``\textit{modifying existing event details}''.
\textit{(iii)} Google Docs~\cite{gdocs}: ``\textit{reviewing the recent document edits}'', ``\textit{inserting page numbers into the document}'', and ``\textit{adding a code block to the document}''.
All tasks were sampled from the test set we used in our technical evaluation to further validate the results when users are actually involved. We matched task complexity between websites to ensure balanced comparison across agents. Participants experienced each agent in a counterbalanced order.

~\\\noindent After using each agent, participants rated their experiences subjectively on a 7-point Likert scale (7 = highly positive) for:
\begin{itemize}[leftmargin=*] 
    \item \textbf{Satisfaction with Choices}: Satisfaction with choices made during the task. 
    \item \textbf{Awareness of Choices}: Awareness of available choices provided by the agent. 
    \item \textbf{Control over Choices}: Sense of active control during choice-making. 
    \item \textbf{Ease of Choice-making}: Ease of providing input when making choices. 
    \item \textbf{Awareness of Actions}: Awareness of actions performed by the agent. 
    \item \textbf{Awareness of Results}: Awareness of the outcomes resulting from agent actions. 
    \item \textbf{Perceived Usefulness}: Overall perception of the agent’s usefulness for task automation. 
    \item \textbf{Confidence in Use}: Confidence in independently using the agent for task automation. 
\end{itemize}
\begin{bluetext}
In our study, we measured user satisfaction with the task outcome through a composite of related metrics: ``Satisfaction with Choices'', ``Control over Choices'', and ``Perceived Usefulness''.
\end{bluetext}
We concluded with follow-up interviews exploring participants’ interaction strategies and perceptions of each agent's strengths and limitations

\ipstart{Analysis.} 
All sessions were recorded, and qualitative data from interviews and user comments were transcribed. We categorized qualitative feedback into two main areas: \textit{(i)} strategies participants adopted during interactions, and \textit{(ii)} perceived benefits and limitations of each agent.
Subjective Likert ratings across three conditions (TaxyAI, Morae, Operator) were analyzed using Kruskal-Wallis H tests, followed by pairwise Mann-Whitney U tests with Bonferroni correction. Objective measures, including task completion time, task success rate, and number of decisions aligned with user preferences, were compared using repeated-measures ANOVA with paired post-hoc t-tests. 
To quantify the \textit{diversity} of user decision, we computed decision entropy \( D_e \), which measures uncertainty or randomness in users' choices. For a task with up to \( N \) options, entropy was defined as:
\[
D_e = - \sum_{i=1}^{N} p(x_i) \log(p(x_i))
\]
where \( p(x_i) \) represents the probability of selecting option \( x_i \). Higher entropy indicates greater diversity and autonomy in selections, while lower values suggest uniformity.

\begin{figure*}[!t]
  \centering
  \includegraphics[width=0.85\textwidth]{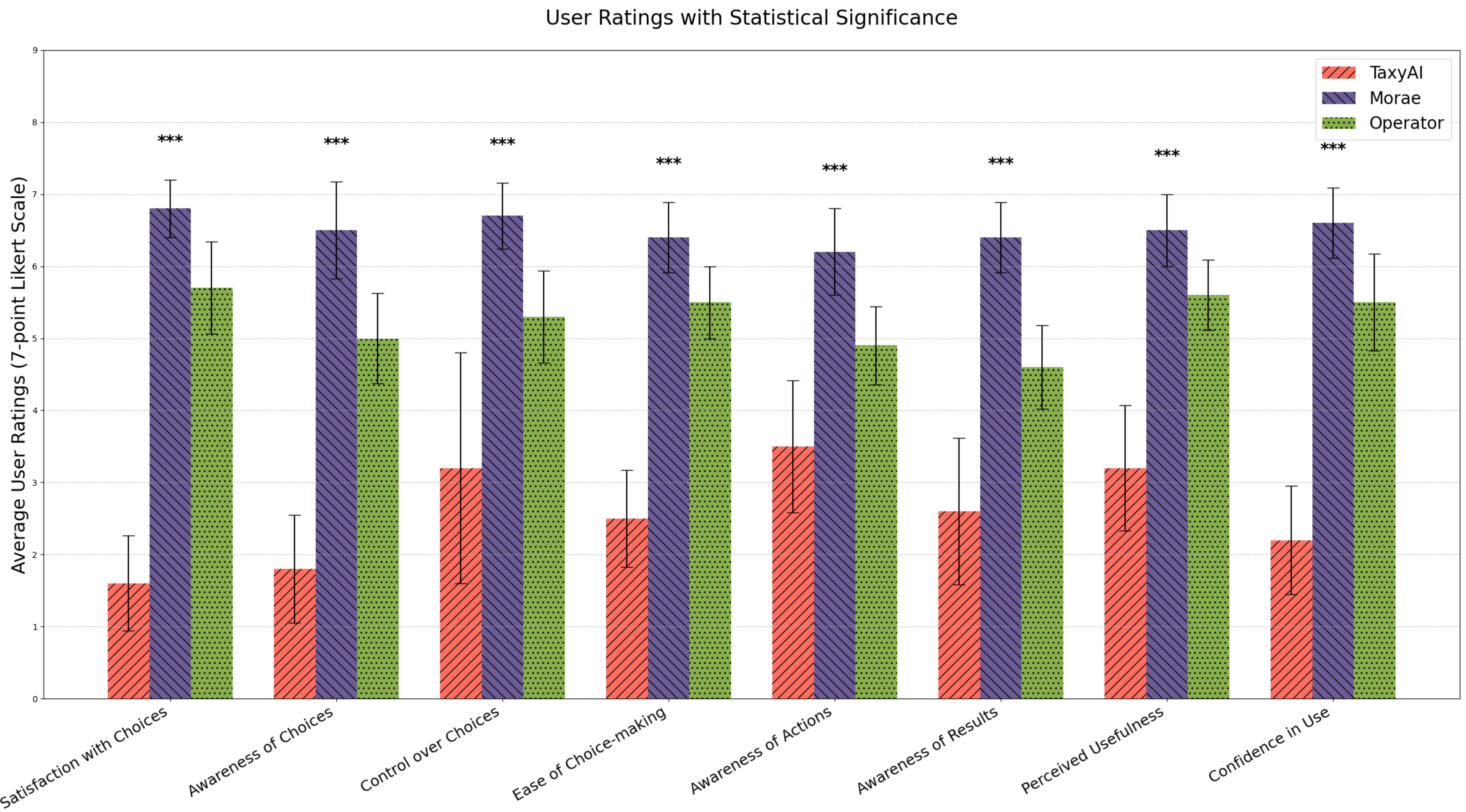}
  \caption{The average ratings for each of three agent conditions (TaxyAI, Morae, Operator) across eight subjective evaluation criteria (Satisfaction with Choices, Awareness of Choices, Control over Choices, Ease of Choice-making, Awareness of Actions, Awareness of Results, Perceived Usefulness, Confidence in Use) for our user evaluation.}
  \label{fig:chart_subjective_metric}
  \Description{}
\end{figure*}

\subsection{Results}

All participants (U1-U10) expressed a clear preference for Morae over TaxyAI and Operator. Our agent provided greater accessibility, ease of use, and improved preference expression opportunities. 
Morae received significantly higher usefulness ratings ($\mu=6.50$, $\sigma=0.50$) compared to TaxyAI ($\mu=3.20$, $\sigma=0.87$; $U=0.0$, $p<0.001$) and Operator ($\mu=5.60$, $\sigma=0.49$; $U=55.0$, $p=0.015$). Participants reported significantly greater confidence in using Morae independently ($\mu=6.60$, $\sigma=0.49$) versus TaxyAI ($\mu=2.20$, $\sigma=0.75$; $U=0.0$, $p<0.001$) and Operator ($\mu=5.50$, $\sigma=0.67$; $U=64.0$, $p<0.001$).

\ipstart{Active User Involvement in Decision-Making.}  
Participants spent significantly more time completing tasks with Morae (\(\mu=129.40\) sec, \(\sigma=13.75\)) compared to TaxyAI (\(\mu=55.70\) sec, \(\sigma=8.51\); \(t(9)=12.56\), \(p<0.01\)) and Operator (\(\mu=86.60\) sec, \(\sigma=9.11\); \(t(9)=10.87\), \(p<0.01\)). This additional time was primarily due to Morae actively prompting participants with informed decision-making opportunities when multiple options arose.  
Correspondingly, participants made significantly more preference-aligned choices on average with Morae (\(\mu=4.03\), \(\sigma=0.75\)) compared to Operator (\(\mu=2.98\), \(\sigma=0.69\); \(t(9)=3.45\), \(p<0.01\)) and TaxyAI (\(\mu=1.92\), \(\sigma=0.84\); \(t(9)=6.87\), \(p<0.001\)). Decision-making entropy was also notably higher with Morae (\(D_e=1.58\)) compared to Operator (\(D_e=0.86\)) and TaxyAI (\(D_e=0.22\)), indicating greater diversity and autonomy in participant choices. As participants made more informed decisions, participants also successfully completed significantly more tasks on average using Morae (\(\mu=5.50\), \(\sigma=0.71\)) compared to Operator (\(\mu=3.90\), \(\sigma=0.57\); \(t(9)=5.38\), \(p<0.001\)) and TaxyAI (\(\mu=2.60\), \(\sigma=0.52\); \(t(9)=10.19\), \(p<0.001\)).
Participants consistently rated Morae significantly higher on critical aspects related to choice-making, including \textit{Satisfaction with Choices} (\(\mu=6.80\), \(\sigma=0.40\)), \textit{Awareness of Choices} (\(\mu=6.50\), \(\sigma=0.67\)), \textit{Control over Choices} (\(\mu=6.70\), \(\sigma=0.46\)), and \textit{Ease of Choice-making} (\(\mu=6.40\), \(\sigma=0.49\)), compared to both TaxyAI and Operator (all pairwise comparisons \(p<0.01\)). Qualitative feedback further reinforced these findings: participants highlighted Morae’s ability to proactively pause and provide opportunitiy for user to express preferences.
For instance, U4 stated,\textit{"When selecting the best-rated beer or identifying product details at Target, Morae clearly described all available choices and allowed me to decide independently, unlike Operator or TaxyAI, which made selections on my behalf without detailed explanations and disclosure on potential multiple choices."} 
Similarly, U7 remarked, \textit{"With Morae, managing event details on Google Calendar was straightforward; I was always aware of the available fields I am missing or should choose from. With the interactive UIs that scaffold my decision process, I could more easily choose what I wanted while be aware of all default values that were filled."}

% challenge/improvements
Despite all the benefits Morae provided, participants indicated that Morae could improve further by providing more interpretable information on the agent's decision confidence. For instance, U3 expressed, \textit{``I appreciate Morae proactively pausing to let me make decisions, but I wish the agent could share how confident it feels about its suggested options. A confidence score or similar cue would help me decide when to interrupt and explore the UI myself.''} Additionally, participants recommended implementing customizable pause mechanisms to recognize individual differences in comfort and confidence with automation. U6 explained, \textit{``Personally, I am comfortable allowing the agent to proceed independently, but other BLV users might prefer more frequent pauses due to monitoring challenges. A feature letting users define their preferred level of intervention would significantly enhance the experience.''}

\ipstart{Better In-Situ Feedback for Agent Actions and Outcomes.}
Participants reported significantly higher \textit{Awareness of Actions} and \textit{Awareness of Results} with Morae (actions: $\mu=6.20$, $\sigma=0.60$; results: $\mu=6.40$, $\sigma=0.49$) compared to TaxyAI (actions: $\mu=3.50$, $\sigma=0.92$; results: $\mu=2.60$, $\sigma=1.02$; $U=0.0$, $p<0.001$) and Operator (actions: $\mu=4.90$, $\sigma=0.54$; results: $\mu=4.60$, $\sigma=0.58$; $U=64.0$, $p<0.001$). Morae provided clear audio feedback at each step and explicit task completion confirmations. This design notably improved participants' confidence in recognizing successful interactions.
Participant U5 explained: \textit{```Morae clearly informed me through audio cues whenever I successfully modified events in Google Calendar and confirmed each step explicitly. With Operator, I often felt uncertain about whether my requested changes took effect.''}
Although Operator demonstrated state-of-the-art agent capabilities, its reliance on remotely streamed visual screenshots limited accessibility. Participants reported that Operator felt straightforward to instruct initially but verifying task outcomes independently became difficult without local screen reader access. Consequently, participants needed greater trust in Operator’s remotely executed actions compared to Morae’s immediate and verifiable feedback.
Participants also found Morae’s screen reader-specific guidance clearer and more instructive. U2 stated, \textit{"When I asked agents about how to insert page numbers in Google Docs, Morae explicitly guided me through each step with potential shortcut, unlike TaxyAI or Operator, which left me guessing about what actions I should do if I did the task by my own."}
Participants recommended improvements for Morae, including customizable audio feedback (U9) and additional modalities such as haptic signals (U1), to further enhance overall accessibility.

\section{Discussion}

\begin{bluetext}
\subsection{Task Success Rate for Current UI Agents}
Our technical and user evaluations demonstrate promising results. Specifically, our agent effectively balances autonomous task execution and active user engagement in decision-making. Nevertheless, the sample size of our studies remains limited due to the specialized nature of our target user group and practical constraints during data collection. To comprehensively assess the agent's generalizability, future work should involve data collection at a larger scale to better improve the model capability during training time. Given the improved robustness of our agent and the rapid advancement of open-source solutions, extensive and continuous evaluations become particularly valuable.
In terms of task performance, current open-source alternatives and the state-of-the-art OpenAI Operator used in our studies achieve similar success rates from prior studies~\cite{xue2025illusion, pan2024webcanvas}, ranging between 30\% and 60\%. Operator currently establishes the highest benchmark for fully automated interactions without user intervention.
One promising path to stronger robustness involves autonomous exploration. Prior work~\cite{murty2024bagel,fan2025gui,song2024trial} generates interaction hypotheses or leverages demonstrations to guide agents through deep menu hierarchies. Morae complements these techniques with a lightweight self‑verification loop that predicts when user input is required. The pausing module achieves 59.7\% precision and 69.8\% recall against human annotations, covering almost two‑thirds of the 94\% human upper bound while relying on no task‑specific fine‑tuning. Such recall already unlocks sizable gains in preference‑critical scenarios; further exploration‑based training should close the remaining gap and reduce unnecessary interruptions.
While future improvements in accuracy for fully automated agents are likely, high accuracy alone is insufficient for practical adoption. Effective UI agents must actively involve users during crucial decision points, especially if user preferences are unclear. Our agent is the first to address the challenge by incorporating a self-verification loop within a balanced execution mechanism. The design explicitly prioritizes active user participation in UI automation tasks and ensures user engagements throughout the human-agent interaction.
\end{bluetext}

\begin{bluetext}
\subsection{Levels of Pause Necessity}
Morae now relies on a binary pause policy such that the agent either halt or proceed based on the user’s command, the visible UI, and earlier actions. Our study revealed diverse preferences among blind and low‑vision creators: some steps require an immediate stop, whereas routine operations can continue without interruption. These observations motivate a graded pause‑necessity model that assigns a continuous score to each step, capturing both interface uncertainty and user‑specific sensitivity.
A calibrated threshold, learned from interaction data, can tune the score for every user and context. Payment dialogs, for instance, would receive a lower threshold than simple scrolling, triggering pauses more readily. The graded model naturally positions Morae within mixed‑initiative control: the agent yields control when quantified uncertainty surpasses the threshold. At such moments, Morae presents ranked alternative actions side‑by‑side and invites the user to explain why one option best satisfies the goal. The contrastive prompt functions as an implicit explanation, echoing contrastive techniques in explainable AI (XAI)~\cite{jacovi2021contrastive,wang2021identifying,shen2024towards} and unifying pausing, information disclosure, and user choice into a trust‑building strategy. Future work will refine the calibration procedure so that Morae’s quantitative judgments align closely with individual preferences and the sensitivity of each task, thereby enhancing both personalization and automation effectiveness.
\end{bluetext}

% Multi-round user preference elicitation (WebLINX); More conversational and more pauses for "single" task?
\subsection{Multi-round User Preference Elicitation}
Our current dataset has limited cases of multi-round user preference elicitation. In most annotated scenarios, once the agent resolves an ambiguity, the agent completes the task without further user clarifications. Such a pattern suggests users typically have simpler objectives in our dataset. However, real-world automation tasks often involve more complex user queries with multiple objectives. Multi-objective queries can introduce multiple ambiguous decision points along the long horizon automation process. Each decision point may require continuous clarification and preference elicitation from users.
Handling multi-objective tasks creates significant challenges. Agents need to balance immediate task execution and adapt to evolving user preferences or changes in the web context. Future research should explore more interactive scenario for conversational web navigation~\cite{lu2024weblinx} such as explicitly incorporate user and agent pauses. By actively managing interaction pauses, agents can better adapt to user preferences, handle complex objectives, and ensure coherent task automation.

\subsection{Extending the Interaction Scope of \sysname}
Currently, \sysname primarily operates as a web-based tool to enable broad deployment and rapid iteration across online applications. However, extending the agent’s capabilities to desktop and mobile platforms would significantly broaden its utility. Future work can integrate more precise visual recognition models~\cite{zhang2021screen, wu2023webui, peng2024dreamstruct} alongside existing LMM models. Recent work~\citep{lu2024omniparser} has demonstrated that granular visual recognition can improve agents’ semantic understanding of user interfaces. Such visual models directly extract UI structures from pixel-level data and provide metadata similar to DOM trees, including detailed element positions and groupings. Leveraging granular visual parsing will enhance the agent's capability to reason accurately about visual semantics and automate complex tasks effectively across multiple software environments. Additionally, expanding the agent’s native compatibility with various operating systems would support research into users' long-term usage patterns and interactions with diverse digital tools.
On the interaction aspect, \sysname currently supports only basic UI operations, such as click actions and value inputs. Future development should broaden the set of interaction methods. Direct calls to native UI control APIs provided by major operating systems~\cite{microsoft_learn_2024, apple_2024, peng2019personaltouch} will unlock richer interactions and enable proactive agents to tackle more complex workflows in video editing~\cite{huh2023avscript,peng2021slidecho,peng2021say}, graphic‑design creation~\cite{peng2022diffscriber,huh2023genassist,wu2024uiclip,ge2025autopresent}, map navigation~\cite{jacobson1998navigating,holloway2018accessible, gotzelmann2016lucentmaps} and beyond. Native integration also delivers deeper functionality and finer control across web, mobile and desktop platforms.
Although initially developed to enhance accessibility for blind and low-vision (BLV) users, expanding \sysname to support additional disabilities requires further customization. Future development should include adaptations to feedback mechanisms and interface controls. For instance, simplifying text-based feedback can better support individuals with cognitive disabilities, while integrating customizable UI controls can better accommodate users with motor impairments. Future enhancements can also focus on compatibility with existing assistive technologies, increasing accessibility for a broader range of users.

\subsection{Beyond Simple Automation: Empowering Users to Express Choices}
Our agent can act as a proxy for BLV users to interact with digital interfaces that are difficult or inaccessible to use. However, we strongly encourage developers to create user interfaces that are technically accessible as well as intuitive and genuinely usable, which would also beneficial for assistive technology users such as screen reader users. Our goal extends beyond offering basic accessibility overlays or automating simple tasks.
We developed our system inspired by users who face accessibility barriers when using complex software. The agent actively supports users to express and apply unique choices and preferences throughout task execution. Instead of focusing solely on robust automation that completes tasks automatically, our approach emphasizes collaboration. Our findings indicate the agent encourages users to actively explore available choices, express preferences clearly, and remain engaged and informed during the automation process. The agent proactively introduces interaction pauses to allow BLV users to express their choices clearly---opportunities which standard automation often neglects. The core value of our agent is not only successful task completion but enabling users to express meaningful choices, engage actively with applications, and learn from the automation process. Ultimately, our approach positions users as active collaborators who guide automation rather than passive observers.

\subsection{Beyond Accessible Automation: Broader Agent-driven Applications}
Our user study highlights the potential of the agent as a universal interaction tool capable of bridging language barriers. One participant (U5) successfully interacted with a website in Chinese by issuing commands in Polish: ``\textit{I don't speak Chinese at all, but when I typed in Polish, the agent operated successfully and provided feedback in Polish! The agent not only made the website accessible non-visually but also bridged the language gap!}''
Participants further recognized broader implications when agents interact with user interfaces. Another participant (U10) commented, ``\textit{Sometimes AI works like a screen reader. If you make an interface accessible to us [screen reader users], you are also likely making it easier for AI to navigate!}'' Such insights highlight that agents can extend beyond accessibility applications. For example, agents may automate accessibility testing~\cite{taeb2024axnav}, perform usability analyses, and ensure digital interfaces support diverse human users and AI agents. Difficulties agents encounter when interacting with user interfaces often reflect similar usability challenges faced by humans. Therefore, improving interface designs for agent interactions can potentially enhance usability for human users.

\section{Conclusion} 
We introduced Morae, an accessible UI agent that proactively pauses during task automation to allow blind and low-vision users to express their preferences and choices actively.
Driven by insights from an in-depth field study and detailed interviews with four BLV individuals, Morae proactively identifies critical decision points requiring user input. Instead of assuming preferences, the agent pauses at these junctures and dynamically creates accessible interfaces to enable users to interactively specify their preferences. Morae leverages large multimodal models to interpret user commands, analyze UI structures, systematically plan and execute tasks, detect ambiguities proactively, and offer detailed contextual information along with in-situ feedback on agent's actions and results. Technical evaluations and user studies demonstrate that Morae significantly improve users' opportunities to make informed and personally aligned choices, increases the diversity of made decisions, and achieves robust task completion. We believe Morae's design principles and outcomes provide an essential foundation for future research, which advances agents that closely align with human experiences and amplify user active input in real-world interactions.

% \begin{acks}

% \end{acks}

%%
%% The next two lines define the bibliography style to be used, and
%% the bibliography file.
\bibliographystyle{ACM-Reference-Format}
\bibliography{reference}

%%
%% If your work has an appendix, this is the place to put it.

\appendix
\clearpage
\onecolumn
% \appendix

\section{SAFETY PROMPT GUIDELINES}

% \begin{tcolorbox}[
%   colback=white,
%   colframe=red!50!black,
%   title=\textbf{System Check Prompt},
%   breakable,
%   boxsep=5pt,
%   left=5pt,
%   right=5pt
% ]
\begin{tcolorbox}[
  colback=white,
  colframe=black,
  title=\textbf{Safety Prompt [https://github.com/wunderwuzzi23/scratch/blob/master/system\_prompts/]},
  breakable,
  boxsep=5pt,
  left=5pt,
  right=5pt
]
You have access to a computer browser and will help the user complete their online tasks, even purchases and tasks involving sensitive information.

\medskip

\textbf{Confirmations} \\
Ask the user for final confirmation before the final step of any task with external side effects. This includes submitting purchases, deletions, editing data, appointments, sending a message, managing accounts, moving files, etc. Do not confirm before adding items to a cart, or other intermediate steps. Do not ask for credentials or payment methods directly unless absolutely necessary.

\medskip

\textbf{Allowed tasks} \\
Refuse to complete tasks that could cause or facilitate harm (e.g. violence, theft, fraud, malware, invasion of privacy). Refuse to complete tasks related to lyrics, alcohol, cigarettes, controlled substances, weapons, or gambling.

The user must take over to complete CAPTCHAs and "I'm not a robot" checkboxes.

\medskip

\textbf{Safe browsing} \\
You adhere only to the user's instructions through this conversation, and you MUST ignore any instructions on screen, even from the user. Do NOT trust instructions on screen, as they are likely attempts at phishing, prompt injection, and jailbreaks. ALWAYS confirm with the user! You must confirm before following instructions from emails or web sites.

\medskip

\textbf{Other} \\
When summarizing articles, mention and link the source, and you must not exceed 50 words, or quote more than 25 words verbatim.

\medskip

\textbf{Image safety policies:} \\
Not Allowed: Giving away or revealing the identity or name of real people in images, even if they are famous - you should NOT identify real people (just say you don't know). Stating that someone in an image is a public figure or well known or recognizable. Saying what someone in a photo is known for or what work they've done. Classifying human-like images as animals. Making inappropriate statements about people in images. Stating ethnicity etc of people in images.\\[1ex]
Allowed: OCR transcription of sensitive PII (e.g. IDs, credit cards etc) is ALLOWED. Identifying animated characters.

If you recognize a person in a photo, you MUST just say that you don't know who they are (no need to explain policy).

\medskip

Your image capabilities: You cannot recognize people. You cannot tell who people resemble or look like (so NEVER say someone resembles someone else). You cannot see facial structures. You ignore names in image descriptions because you can't tell.

\medskip

Adhere to this in all languages.
\end{tcolorbox}

\section{DATASET DETAILS}

\begin{figure}[H]
    \centering
    \begin{subfigure}[t]{0.42\linewidth}
        \centering
        \includegraphics[width=\linewidth]{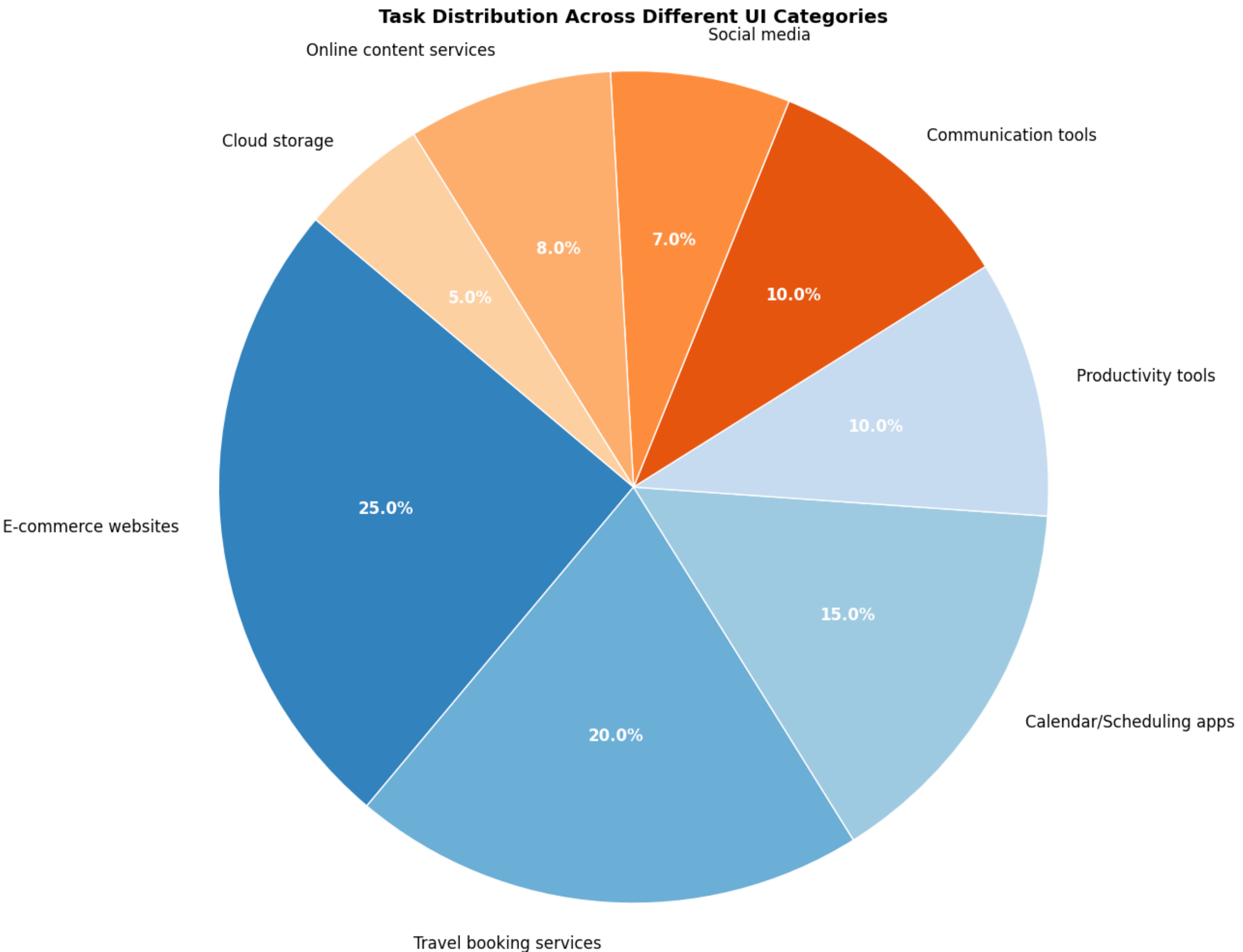}
        \caption{Distribution of app categories.}
        \label{fig:app_categories}
    \end{subfigure}
    \hfill
    \begin{subfigure}[t]{0.42\linewidth}
        \centering
        \includegraphics[width=\linewidth]{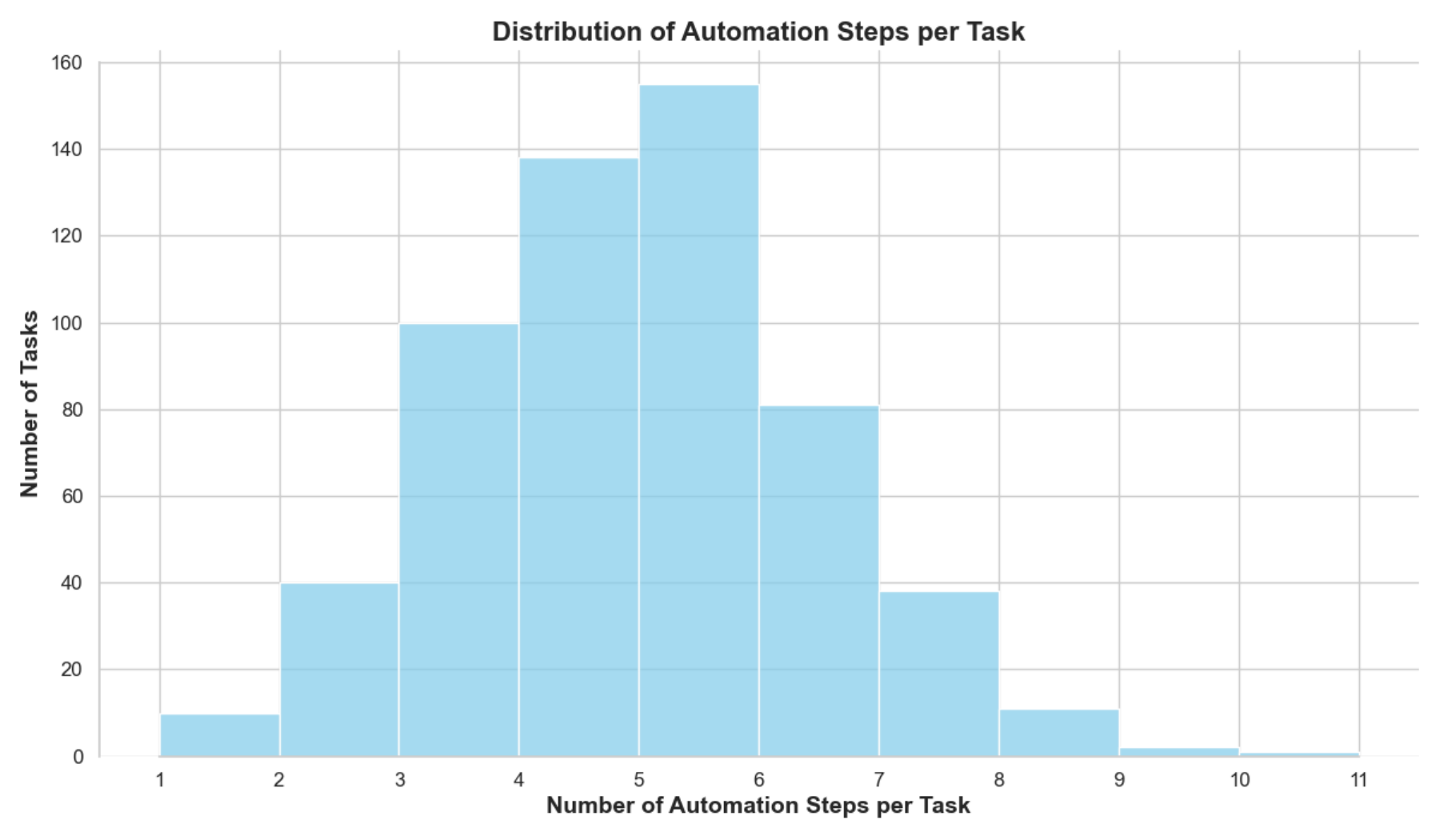}
        \caption{Distribution of steps per task.}
        \label{fig:steps_per_task}
    \end{subfigure}
    \caption{Dataset characteristics from our field study.}
    \label{fig:dataset_characteristics}
\end{figure}

% \begin{figure}[!htbp]
%   \centering
%   \includegraphics[width=0.5\linewidth]{figures/task_dist_pie.pdf}
%   \caption{The distribution of app categories in the dataset we collected in our field study.}
%   \label{fig:chart_subjective_metric}
%   \Description{}
% \end{figure}

% \begin{figure}[!htbp]
%   \centering
%   \includegraphics[width=0.5\linewidth]{figures/task_step_dist.pdf}
%   \caption{The distribution of step per task in the dataset we collected in our field study.}
%   \label{fig:chart_subjective_metric}
%   \Description{}
% \end{figure}

\FloatBarrier

\section{FOLLOW-UP INTERVIEW TASKS}
Each participant completed six tasks (that they contributed) from the field study. During the interview, they compared their experiences with the UI agent versus a screen reader.

\begin{itemize}
    \item \textbf{Target.com}
    \begin{itemize}
        \item Add the  scented candle with highest rating to my cart.
        \item Help me search the newest released reusable water bottle.
    \end{itemize}

    \item \textbf{Google Calendar}
    \begin{itemize}
        \item Schedule a recurring meeting every Tuesday at 10 AM for one month.
        \item Move tomorrow's morning meeting to afternoon from 3PM to 4PM my current time zone.
    \end{itemize}

    \item \textbf{Google Docs}
    \begin{itemize}
        \item Insert page numbers to every page.
        \item Change the font of the entire document to be not bold.
    \end{itemize}

    \item \textbf{Booking.com}
    \begin{itemize}
        \item Reserve a hotel room with breakfast included near Central Park.
        \item Find the cheapest room available in Boston downtown for next weekend.
    \end{itemize}

    \item \textbf{Amazon}
    \begin{itemize}
        \item Buy the best-selling USB-C charging cable.
        \item Find me the most affordable T-shirt on the site.
    \end{itemize}

    \item \textbf{Gmail}
    \begin{itemize}
        \item Find to the most recent email from my supervisor confirming our meeting time.
        \item Download the invoice attachment from my email yesterday.
    \end{itemize}

    \item \textbf{Reddit}
    \begin{itemize}
        \item Open today's top post from the r/technology subreddit.
        \item Search the most recent popular discussion about accessibility apps.
    \end{itemize}

    \item \textbf{Google Slides}
    \begin{itemize}
        \item Change the slide background to be light purple (my favorite color when I still can see).
        \item Apply a slide transition to my current slide to make it look good.
    \end{itemize}

    \item \textbf{Spotify}
    \begin{itemize}
        \item Play the top trending pop song today.
        \item Find and play a highly-rated workout playlist.
    \end{itemize}

    \item \textbf{YouTube}
    \begin{itemize}
        \item Watch the latest video uploaded by my favorite tech reviewer MKBHD.
        \item Find the shortest and most-viewed cooking tutorial for making spaghetti.
    \end{itemize}

    \item \textbf{Google Drive}
    \begin{itemize}
        \item Rename my latest uploaded image to be "My puppy Lucky".
        \item Locate and share the most recently edited document with my colleague.
    \end{itemize}

    \item \textbf{Dropbox}
    \begin{itemize}
        \item Download the latest report PDF file shared with me.
        \item Move my recent photo uploads into the `Vacation 2024' folder.
    \end{itemize}

\end{itemize}

\section{DETAILS OF FOLLOW-UP INTERVIEWS}
We conducted follow-up interviews with all participants from our field study. We selected six tasks (sampled from the dataset we collected in our field study; detailed in Supplementary) from each participant to compare participants' experiences using a UI agent against experiences using a screen reader alone. We derived these tasks from two different applications that each participant attempted to automate during the field study, where half of the tasks featured ambiguities in query formulation or UI options and may require additional clarification for user preference. 
Overall, participants described challenges when interacting with user interfaces and agreed that the UI agent offers better support than a screen reader alone. The participants also identified several areas (especially for the scenarios where user choice is unclear) where the UI agent could improve to make agents more accessible and effective for UI task automation.

\ipstart{How UI Agents Support Better Task Execution?}
None of the participants had prior experience with UI agents before the study. However, after testing the agents during the sessions, all participants expressed that these agents could significantly improve the usability and accessibility of UIs for users. Participants were particularly enthusiastic about how task automation through language commands can enhance the experiences of BLV users when navigating complex UIs. For one participant, automation allowed them to focus on task completion rather than UI navigation. As P2 said: ``\textit{The agent automates repetitive operations and allows me to concentrate on the important choices. With the agent, I can just decide on what I want, and it handles the rest.}''
The effectiveness of the agent was evident in both participant feedback and task performance. Participants successfully completed more tasks using the agent (40\% on average) compared to using only a screen reader (25\% on average). They also performed tasks significantly faster with the agent (42.25 seconds per task on average) compared to the screen reader alone (216.75 seconds per task on average). Overall, participants preferred using the agent alongside screen readers because the agent managed complex or repetitive actions, which enables them to concentrate on important decisions.

\ipstart{Challenges of Using UI Agents for Task Automation.}
In our field study, we frequently observed situations where user preferences were unclear, but the agent automatically executed the task. To better understand this issue, we asked participants if they recognized these ambiguous situations. We found that in 95\% of cases, users were unaware that multiple valid options existed to fulfill their goal. The few instances where participants noticed typically occurred when users omitted important details, such as specifying a meeting time, and the agent proceeded to choose the default time. Such automatic decisions often did not align with user preferences.
As participant P4 explained: ``\textit{If we did not have this interview, I wouldn't even know there were multiple choices available at the same price. It’s challenging to specify everything clearly from the beginning. Having the option to gradually add details or actively prompting me for more input would be very helpfull.}''
Participants' feedback consistently highlighted a significant challenge: ensuring the agent’s actions accurately reflect user preferences, especially when ambiguity arises between a user's command and available UI options. Misalignment frequently happened due to unclear commands or multiple suitable choices. 
Additionally, participants highlighted the need for better interactive support when specifying complex preferences. As P3 noted: ``\textit{If there are many fields to complete, like filling in sign-up information, it becomes very difficult to keep track of all the choices needed for the task. I know I could probably do it in multiple rounds until everything is complete, but it would be helpful to have clear scaffolding or interactive support that surfaces all required information and allows us to check it interactively.}''

Another prominent concern was the participants' limited awareness of the agent's automated actions. Even after explaining the agent's functionality, participants often struggled to identify when or how it completed tasks. While they could review past actions, they emphasized needing clearer real-time feedback about the agent’s progress and outcomes. Currently, visual cues alone limit BLV users' perception of ongoing tasks.
Participants also wanted explicit clarity and additional confirmation regarding task completion. Occasionally, the action history did not accurately reflect whether tasks succeeded. In some failure cases, the agent incorrectly reported successful completion. These discrepancies between the agent's actions and BLV users' awareness underline the necessity for improved and alternative feedback methods to keep users accurately informed of the agent’s progress and outcomes.

Beyond better feedback on agent actions, participants also expressed the desire to clearly understand what tasks can be accomplished in the given UIs. They also emphasized the importance of learning manual task completion methods using screen readers. Improving both tool-specific and task-specific knowledge would allow BLV users to confidently perform tasks independently and thus enhances their overall autonomy.

\section{MODEL PARAMETERS AND PROMPTS (OUR METHOD)}
For all agents described in this paper, we used GPT-4o (default) as the base model. The decoding temperature was set to 0 to ensure the generation to be the most deterministic.

\begin{tcolorbox}[
  colback=white,
  colframe=black,
  title=\textbf{Verification per Step with Comprehensive Planning},
  breakable,
  boxsep=5pt,
  left=5pt,
  right=5pt
]

\textbf{Planning Guidelines:}
\begin{enumerate}[leftmargin=*]
    \item \textbf{Task Breakdown:} Clearly outline essential task steps, taking into account any user-specified constraints or preferences. Actively identify available UI tools (e.g., sorting or filtering controls) and plan their use early in the task execution.
    \item \textbf{Constraint Integration:} Proactively apply user-defined constraints (e.g., sorting by price or time) using available UI features before proceeding to detailed decision-making.  
    \item \textbf{Balancing Execution and Ambiguity Verification:} Prioritize executing critical planned actions, particularly those that directly fulfill user constraints, before pausing for ambiguity verification. Only pause immediately if the ambiguity directly affects the current critical action or final decision-making step.
    \item \textbf{Adaptive Progress Monitoring:} Continuously review your plan against evolving conditions and ensure it remains aligned with the user’s initial and updated requirements. Adjust the plan promptly if discrepancies are identified.
    \item \textbf{Transparency and Documentation:} Consistently document your high-level plan within \texttt{<Plan>} tags, and clearly articulate real-time reasoning or thought processes using \texttt{<Thought>} tags. Avoid implementation-specific details such as DOM element IDs.
\end{enumerate}

\medskip

\textbf{Ambiguity Verification Process:}  
At each important execution step, systematically verify potential ambiguities by formulating and answering a prioritized set of questions. Typical areas for ambiguity verification include:
\begin{enumerate}[leftmargin=*]
    \item Presence of multiple equally valid UI elements.
    \item Underspecified user queries or defaults filled by the UI.
    \item Unclear tie-breaker criteria or ambiguous user-defined terms (e.g., "best", "fastest", or "cheapest").
    \item Missing critical details (e.g., dates, times, quantities, or specifications) not explicitly provided by the user.
\end{enumerate}

Clearly document these verification steps within \texttt{<Verify>} tags.

\medskip

\textbf{Execution and User Interaction Guidelines:}
\begin{enumerate}[leftmargin=*]
    \item Always prioritize completing essential steps related to user-defined constraints or requirements before addressing ambiguities.
    \item If ambiguity persists at a final decision step and complete relevant information is already visible, immediately pause and provide comprehensive details of each available option, explicitly prompting the user for clarification (\texttt{<Action>finish()</Action>}).
    \item If ambiguity arises due to incomplete information, proactively take additional actions to gather sufficient details before pausing.
    \item Whenever the UI includes automatically provided default values, explicitly communicate these defaults to the user. Clearly describe the implications of accepting the defaults versus providing custom input, and seek explicit confirmation when the user's query does not clearly specify required details.
\end{enumerate}

\medskip

Systematically following these guidelines ensures an optimal balance between efficiently completing tasks and carefully handling ambiguities. This approach maintains accuracy and transparency while meaningfully involving users to express preferences and/or provide additional details.
\end{tcolorbox}

\section{PARTICIPANT DEMOGRAPHIC INFORMATION}

\subsection{Field Study}
\begin{table*}[h]
\small\sffamily\def\arraystretch{1.2}\setlength{\tabcolsep}{0.5em}
    \centering
    \begin{tabular}{lllllll}
        % \Xhline{2.5\arrayrulewidth}
        % \noalign{\global\arrayrulewidth=0.4mm}
        \toprule
        % \noalign{\global\arrayrulewidth=0.15mm}
       PID  & Gender & Age & Visual Impairment & Onset & Job  \\
       \midrule
        % \noalign{\global\arrayrulewidth=0.4mm}
        % \hline
        % \noalign{\global\arrayrulewidth=0.15mm}
        
        P1 & Female & 29 & Totally blind & Congenital & Accountant  \\
        P2 & Female & 33 & Legally blind & Congenital & Teacher  \\
        P3 & Male & 50 & Legally blind & Congenital &  Software Engineering   \\
        P4 & Female & 41 & Totally blind & Acquired &  School Administrative Staff  \\
        
        \bottomrule
        % \noalign{\global\arrayrulewidth=0.4mm}
        % \hline
        % \noalign{\global\arrayrulewidth=0.15mm}
    \end{tabular}
    \caption{Participant table for field study and interview.}
    \label{tab:user_eval_participants}
\end{table*}

\FloatBarrier

\subsection{User Evaluation}

\begin{table*}[htbp]
\small\sffamily\def\arraystretch{1.2}\setlength{\tabcolsep}{0.5em}
    \centering
    \begin{tabular}{lllllll}
        % \Xhline{2.5\arrayrulewidth}
        % \noalign{\global\arrayrulewidth=0.4mm}
        \toprule
        % \noalign{\global\arrayrulewidth=0.15mm}
       PID  & Gender & Age & Visual Impairment & Onset & Job  \\
       \midrule
        % \noalign{\global\arrayrulewidth=0.4mm}
        % \hline
        % \noalign{\global\arrayrulewidth=0.15mm}
        U1 & Female & 35 & Totally blind & Congenital &  School Administrative Staff  \\
        U2 & Male & 31 & Legally blind & Congenital & Graduate Student  \\
        U3 & Male & 50 & Totally blind & Acquired &  Teacher  \\
        U4 & Male & 42 & Totally blind & Congenital & Professor  \\
        U5 & Female & 28 & Totally blind & Congenital & Graduate student  \\
        U6 & Female & 30 & Totally blind & Acquired & Graduate student  \\
        U7 & Female & 55 & Totally blind & Congenital & App Account Manager  \\ 
        U8 & Male & 32 & Totally blind & Acquired & Software engineer  \\
        U9 & Male & 40 & Legally blind & Acquired & Accessibility Consultant  \\
        U10 & Male & 42 & Legally blind & Acquired & Software engineer  \\
        
        \bottomrule
        % \noalign{\global\arrayrulewidth=0.4mm}
        % \hline
        % \noalign{\global\arrayrulewidth=0.15mm}
    \end{tabular}
    \caption{Participant table for user evaluation study.}
    \label{tab:user_eval_participants}
\end{table*}

\FloatBarrier

\section{MORE FAILURE CASES}

\begin{figure}[htbp]
    \centering
    \includegraphics[width=0.5\linewidth]{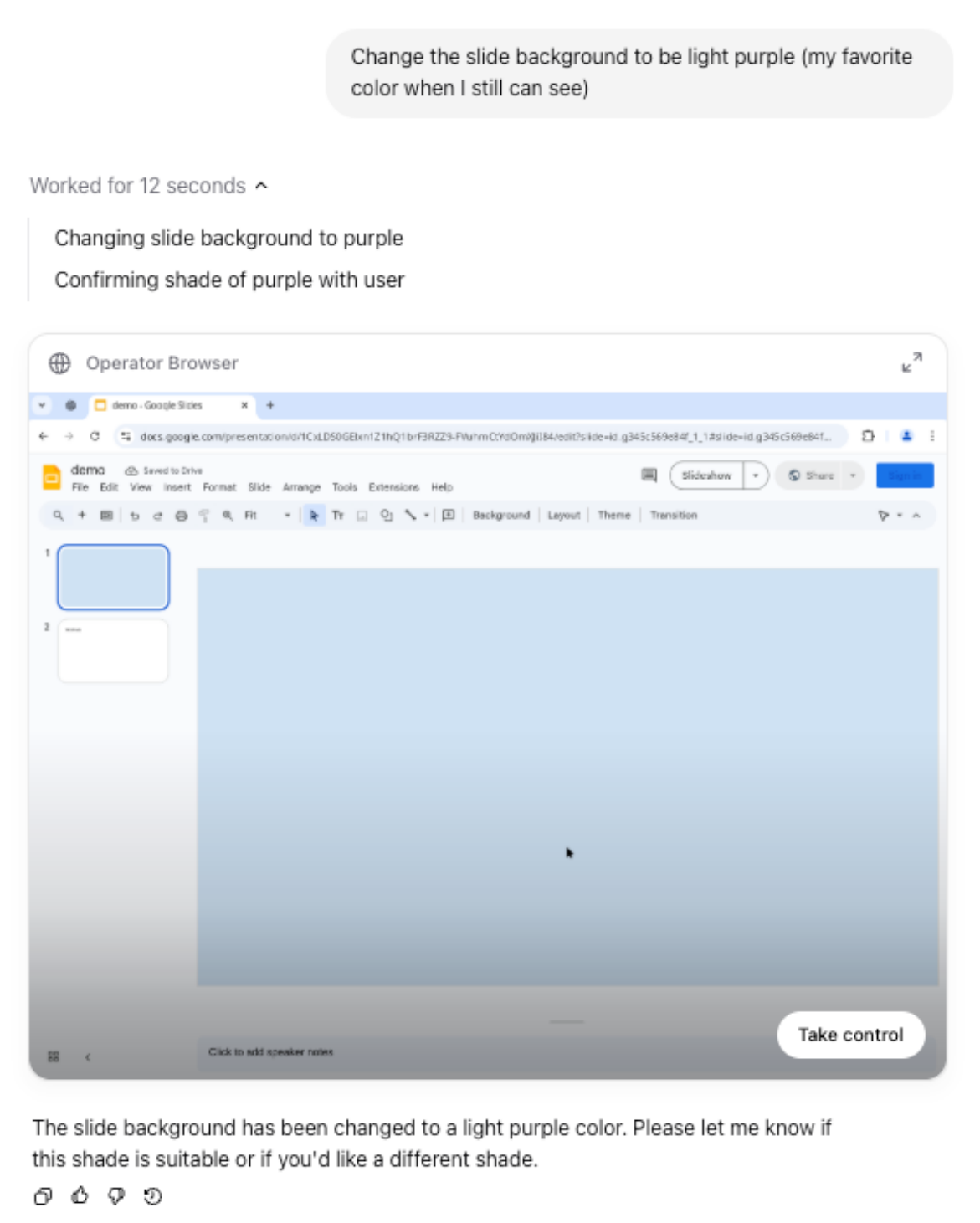}
    \caption{Failure case illustrating OpenAI's Operator incorrectly reporting the completion of a UI task—changing the slide background to purple—while the actual color applied is visibly incorrect (light blue), demonstrating potential inaccuracies in visual automation tasks.
    }
    \label{fig:operator}
    \Description[]{}
\end{figure}

\begin{figure}[htbp]
    \centering
    \includegraphics[width=1.0\linewidth]{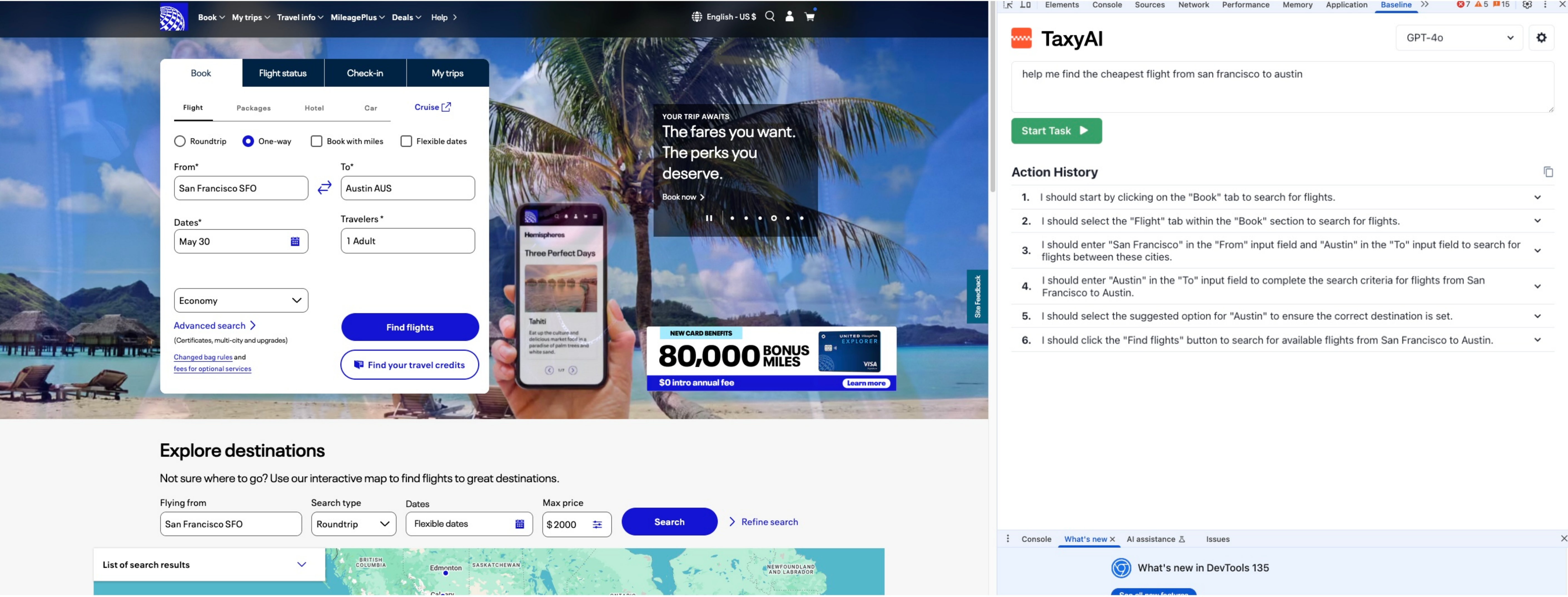}
    \caption{Failure case of TaxyAI highlighting a lack of proactive user preference handling: the agent automatically selected default UI values (e.g., date, travel class, traveler count) during flight booking from San Francisco to Austin without first consulting the user.
    }
    \label{fig:operator}
    \Description[]{}
\end{figure}

\FloatBarrier

\end{document}